\documentclass[11pt,a4paper]{article}
\usepackage{jheppub}
\usepackage{amssymb,amssymb}
\usepackage{lineno}
\usepackage{subfigure}
\usepackage{comment}
\usepackage{morefloats}
\usepackage{hyperref}

\usepackage[compact]{titlesec}
\titlespacing{\section}{0pt}{8pt}{4pt}{\raggedright}
\titlespacing{\subsection}{0pt}{8pt}{4pt}
\titlespacing{\subsubsection}{0pt}{8pt}{4pt}

\newlength{\figwidth}
\notoc

\def\HighestDijetMass{4.0}

\def\integLumi{4.8~fb$^{-1}$}

\def\MjjLimitqstar{2.83} 
\def\MjjLimitExpectedqstar{2.94} 

\def\MjjLimits8{1.86} 
\def\MjjLimitExpecteds8{1.97} 

\def\MjjLimitWprime{1.68} 
\def\MjjLimitExpectedWprime{1.74} 

\def\MjjLimitStrRes{3.61} 
\def\MjjLimitExpectedStrRes{3.47} 

\def\ElevenBinChiExpectedLambda{7.X}

\def\ElevenBinChiQBH{4.11}
\def\ElevenBinChiExpectedQBH{4.20}

\def\ElevenBinChiLambdaDest{7.6}
\def\ElevenBinChiExpectedLambdaDest{7.7}

\def\ElevenBinChiExpectedLambda2SigMinusDest{7.06}

\usepackage{preprintcover}  

\PreprintCoverPaperTitle{ATLAS search for new phenomena in dijet mass and angular distributions
using $pp$ collisions at $\sqrt{s}=7$~TeV}  

\PreprintIdNumber{CERN-PH-EP-2012-257}  

\PreprintCoverAbstract {Mass and angular distributions of dijets produced in
LHC proton-proton collisions at a centre-of-mass energy $\sqrt{s}$ = 7 TeV
have been studied with the ATLAS detector using the full 2011 data set with
an integrated luminosity of \integLumi. Dijet masses up to
$\sim$ 4.0 TeV have been probed. 
No resonance-like features have been observed in the dijet mass spectrum, 
and all angular distributions are consistent with the predictions of QCD.
Exclusion limits on six hypotheses of new phenomena have been set at 95\% CL
in terms of mass or energy scale, as appropriate. These hypotheses include
excited quarks below \MjjLimitqstar~TeV,
colour octet scalars below \MjjLimits8~TeV, 
heavy $W$ bosons below \MjjLimitWprime~TeV, 
string resonances below \MjjLimitStrRes~TeV,
quantum black holes with six extra space-time dimensions
   for quantum gravity scales below \ElevenBinChiQBH~TeV, 
and quark contact interactions below a compositeness scale of 
\ElevenBinChiLambdaDest~TeV in a destructive interference scenario.
}  

\PreprintJournalName{Journal of High Energy Physics}  

\begin{document}



\def\topbottomLabel{~}
\def\leftrightLabel{~}

\def\Pythia{{\sc Pythia}}
\def\BlackMax{BlackMax}
\def\Geant{{\sc Geant4}}
\def\CalcHEP{CalcHEP}
\def\NLOJET{{\sc NLOJET++}}
\def\BumpHunter{{\sc BumpHunter}}
\def\TailHunter{{\sc TailHunter}}

\def\GeVcc{${\rm GeV}$}
\def\TeVcc{${\rm TeV}$}
\def\GeVc{${\rm GeV}$}
\def\invpb{${\rm pb}^{-1}$}
\def\mFchi{F_{\chi}}
\def\Fchi{$\mFchi$}
\def\mFchimjj{F_{\chi}(m_{jj})}
\def\Fchimjj{$\mFchimjj$}
\def\mjj{m_{jj}}
\def\ystar{{y^\ast}}
\def\qstar{{q^\ast}}
\def\Wstar{{W^\ast}}
\def\Wprime{W'}
\def\qbar{$\overline{q}$}
\def\qqbar{$q\overline{q}$}

\def\pT{$p_{\rm T}$}
\def\mpT{p_{\rm T}}
\def\Acc{{\cal A}}   
\def\thetanp{\theta_{np}}
\def\Nev{{N_{ev}}}


\def\FchimjjLimitQstar{2.75}
\def\FchimjjLimitExpectedQstar{2.85}
\def\FchimjjLimitQstarPval{0.XX}
\def\FchimjjLimitQstarPvalNoOffs{0.XX}

\def\FchimjjLimitAxigluon{3.XX} 
\def\FchimjjLimitExpectedAxigluon{3.XX} 

\def\FchimjjLimitQBH{4.03}
\def\FchimjjLimitExpectedQBH{4.16}

\def\FchimjjLimits8{2.XX} 
\def\FchimjjLimitExpecteds8{1.XX} 

\def\FchimjjLimitWstar{1.XX} 
\def\FchimjjLimitExpectedWstar{1.XX} 

\def\FchimjjLimitWprime{1.XX} 
\def\FchimjjLimitExpectedWprime{1.XX} 

\def\FchimjjLimitStrRes{2.30} 
\def\FchimjjLimitExpectedStrRes{2.27} 


\def\FchimjjLimitLambda{6.X}  
\def\FchimjjLimitExpectedLambda{7.X}


\def\FchimjjLimitLambdaDest{7.6}  
\def\FchimjjLimitExpectedLambdaDest{7.7}  
\def\FchimjjLimitExpectedLambda2SigPlusDest{10.1}  
\def\FchimjjLimitExpectedLambdaSigPlusDest{8.2}  
\def\FchimjjLimitExpectedLambdaSigMinusDest{7.0}  
\def\FchimjjLimitExpectedLambda2SigMinusDest{6.8}  


\def\FchimjjLimitLambdaCons{6.X}  
\def\FchimjjLimitExpectedLambdaCons{7.X}  

\def\ElevenBinChiLambdaCons{7.X}
\def\ElevenBinChiExpectedLambdaCons{7.X}
\def\ElevenBinChiExpectedLambdaSigPlusCons{8.X}
\def\ElevenBinChiExpectedLambdaSigMinusCons{7.X}

\title{ATLAS search for new phenomena in dijet mass and angular distributions
using $pp$ collisions at $\sqrt{s}=7$~TeV}


\collaboration{ATLAS Collaboration}


\date{12 March 2012}



\abstract{Mass and angular distributions of dijets produced in
LHC proton-proton collisions at a centre-of-mass energy $\sqrt{s}=7$~TeV
have been studied with the ATLAS detector using the full 2011 data set with
an integrated luminosity of \integLumi. Dijet masses up to
$\sim \HighestDijetMass$~TeV have been probed. 
No resonance-like features have been observed in the dijet mass spectrum, 
and all angular distributions are consistent with the predictions of QCD.
Exclusion limits on six hypotheses of new phenomena have been set at 95\% CL
in terms of mass or energy scale, as appropriate. These hypotheses include
excited quarks below \MjjLimitqstar~TeV,
colour octet scalars below \MjjLimits8~TeV, 
heavy $W$ bosons below \MjjLimitWprime~TeV, 
string resonances below \MjjLimitStrRes~TeV,
quantum black holes with six extra space-time dimensions
   for quantum gravity scales below \ElevenBinChiQBH~TeV, 
and quark contact interactions below a compositeness scale of 
\ElevenBinChiLambdaDest~TeV in a destructive interference scenario.}


\keywords{Dijets, New phenomena, Excited quarks}


\maketitle




\section{Introduction}

At the CERN Large Hadron Collider (LHC), collisions with the largest momentum
transfer typically result in final states with two jets of particles
with high transverse momentum (\pT).
The study of these events tests the Standard Model (SM) at the highest
energies accessible at the LHC.  At these energies,
new particles could be produced~\cite{Baur:1987ga,Baur:1989kv},
new interactions between particles could manifest 
themselves~\cite{Frampton1987a,Frampton1987b,Bagger1988,Han:2010rf},
or interactions resulting from the unification of SM with gravity could
appear in the TeV range~\cite{RandallMeade,Feng:2004,StrRes1,StrRes2,StrRes3,StrRes4}.
These collisions also probe the structure of the fundamental constituents of matter at the
smallest distance scales allowing, for example, an experimental test of the size of quarks. 
The models for new phenomena (NP) tested in the current studies 
are described in section \ref{sec:SimulNP}.

The two jets emerging from the collision may be reconstructed to  
determine the two-jet (dijet) invariant mass, $\mjj$, and the
scattering angular distribution with respect to the colliding beams of protons.
The dominant Quantum Chromodynamics (QCD) interactions for this
high-\pT\ scattering regime are \mbox{$t$-channel} processes, 
leading to angular distributions that peak at small
scattering angles.  Different classes of new phenomena are expected to modify dijet
mass distribution and the dijet angular distributions as a function of $\mjj$, creating either
a deviation from the QCD prediction above some threshold or an excess of
events localised in mass (often referred to as a ``bump'' or ``resonance'').
Most models predict that the angular distribution of the NP signal would be more isotropic
than that of QCD.

Results from previous studies of dijet mass and angular distributions
\cite{Arnison1986244,Bagnaia1984283,CDF:2009DijetSearch,DZero:2009DijetAng,
      ATLAS:2010bc,ATLAS:2010eza,CMS:2010dijetmass,CMS:2010centrality,
      CMS:2011DijetMassAngle,CMS:Res2011,Aad:2011aj,ATLAS:Res2011} 
were consistent with QCD predictions.   
The study reported in this paper is based on $pp$\ collisions at a centre-of-mass (CM)
energy of 7~TeV produced at the LHC and measured by the ATLAS detector. 
The analysed data set corresponds to an integrated luminosity of  
\integLumi\ collected in 2011 \cite{ATLAS-CONF-2011-116,ATLASLumiEJPC2011}, a substantial increase over
previously published ATLAS dijet analyses~\cite{Aad:2011aj,ATLAS:Res2011}. 

A detailed description of the ATLAS detector has been published 
elsewhere~\cite{DetectorPaper}.  The detector is instrumented over almost
the entire solid angle around the $pp$\ collision point with layers of
tracking detectors, calorimeters, and muon chambers.

High-transverse-momentum hadronic jets in the analysis
are measured using a finely-segmented calorimeter system, designed to achieve a 
high reconstruction efficiency and an excellent energy resolution.
The electromagnetic calorimetry is provided by
high-granularity liquid argon (LAr) sampling calorimeters, using lead as an absorber, 
that are split into a barrel ($|\eta|<1.475$\footnote{
In the right-handed ATLAS coordinate system, the pseudorapidity 
$\eta$\ is defined as $\eta \equiv$ $-$ln tan($\theta$/2), where the polar angle $\theta$
is measured with respect to the LHC beamline. The azimuthal angle $\phi$
is measured with respect to the $x$-axis, which points toward the centre
of the LHC ring. The $z$-axis is parallel to the anti-clockwise beam
viewed from above. Transverse momentum and energy are defined as $p_\textrm{T} = p$\,sin$\theta$
and $E_\textrm{T} = E$\,sin$\theta$, respectively.}) and end-cap ($1.375<|\eta|<3.2$) regions. 
The hadronic calorimeter is divided into barrel, extended barrel ($|\eta|<1.7$) 
and Hadronic End-Cap (HEC; $1.5<|\eta|<3.2$) regions. The barrel and extended
barrel are instrumented with scintillator tiles and steel absorbers, 
while the HEC uses copper with liquid argon modules.
The Forward Calorimeter region (FCal; $3.1<|\eta|<4.9$) 
is instrumented with LAr/copper and LAr/tungsten modules
to provide electromagnetic and hadronic energy measurements, respectively. 

\section{Overview of the dijet mass and angular analyses}
\label{sec:overview}
The dijet invariant mass, $\mjj$, is calculated from the vectorial sum of the 
four-momenta of the two highest \pT\ jets in the event.
A search for resonances is performed on the $\mjj$ spectrum, 
employing a data-driven background estimate that does not rely on QCD calculations.

The angular analyses employ ratio observables and normalised
distributions to substantially reduce their sensitivity
to systematic uncertainties, especially those associated
with the jet energy scale (JES), parton distribution functions (PDFs) and the integrated luminosity.
Unlike the $\mjj$ analysis, the angular analyses use a background estimate based on QCD.
The basic angular variables and distributions used in the 
previous ATLAS dijet studies~\cite{ATLAS:2010eza,Aad:2011aj} are
also employed in this analysis.
A convenient variable that emphasises the central scattering region is $\chi$.
If $E$\ is the jet energy and $p_z$\ is the $z$-component of the
jet's momentum, the rapidity of the jet is given by
$y \equiv \frac{1}{2}\mathrm{ln}(\frac{E + p_z}{E - p_z})$. 
In a given event, the rapidities of the two highest \pT\ jets in the
$pp$ centre-of-mass frame are denoted by $y_1$\ and $y_2$, and the rapidities
of the jets in the dijet CM frame are $y^* = \frac{1}{2}(y_1 - y_2)$ and $-y^*$.
The longitudinal motion of the dijet CM system in the $pp$ frame
is described by the rapidity boost, $y_B = \frac{1}{2}(y_1 + y_2)$.
The variable $\chi$ is: $\chi \equiv \mathrm{exp}(|y_1-y_2|)=\mathrm{exp}(2|y^*|)$.

The $\chi$\ distributions predicted by QCD are relatively flat compared to
those produced by new phenomena.
In particular, many NP signals are more
isotropic than QCD, causing them to peak at low values of $\chi$.
For the $\chi$ distributions in the current studies, the rapidity coverage extends to $|y^*|<1.7$\,
corresponding to $\chi < 30.0$. This interval is divided into 11 bins, with 
boundaries at $\chi_i = \mathrm{exp}(0.3 \times i)$ with $i = 0,...,11$, 
where 0.3 corresponds to three times the coarsest calorimeter segmentation, $\Delta\eta$ = 0.1.  
These $\chi$ distributions are measured in five dijet mass ranges with 
the expectation that low $\mjj$ bins will be dominated by QCD processes and
NP signals would be found in higher mass bins.
The distributions are normalised to unit area, restricting the analysis to a shape comparison.
 
To facilitate an alternate approach to the study of dijet angular
distributions, it is useful to define a single-parameter measure of
isotropy as the fraction $F_{\chi} \equiv \frac{N_\mathrm{central}}{N_\mathrm{total}}$, where
$N_\mathrm{total}$ is the number of events containing a dijet that passes all
selection criteria, and $N_\mathrm{central}$ is the subset of these events in which 
the dijet enters a defined central region. It was found that $|y^*|<0.6$, corresponding
to $\chi < 3.32$, defines an optimal central region where many new processes
would be expected to deviate from QCD predictions.  This value
corresponds to the upper boundary of the fourth bin in the $\chi$ 
distribution.

As in previous ATLAS studies~\cite{ATLAS:2010eza}, the current angular analyses make use of
the \Fchimjj\ distribution, which consists of $F_{\chi}$ binned finely
in $\mjj$: 
\begin{equation}
F_{\chi}(m_{jj}) \equiv \frac{\mathrm{d}   N_\mathrm{central}/\mathrm{d}
  m_{jj} } {  \mathrm{d}   N_\mathrm{total}/\mathrm{d} m_{jj}}, 
\label{eq:fchimjjdef}
\end{equation}
using the same mass binning
as the dijet mass analysis. 
This distribution is more sensitive to mass-dependent changes in the rate
of centrally produced dijets than the $\chi$ distributions but is less sensitive
to the detailed angular shape. 
The distribution of \Fchimjj\ in the central region defined above is similar to
the $\mjj$ spectrum, apart from an additional selection criterion on the boost
of the system (as explained in section \ref{sec:recocuts}).

Dijet distributions from collision data are not corrected (unfolded) for detector
resolution effects. Instead, the measured distributions are compared to theoretical
predictions passed through detector simulation.

\section{Jet calibration}
\label{sec:jetCalibration}

The calorimeter cell structure of ATLAS is designed to follow the
shower development of jets.  Jets are reconstructed from topological
clusters (topoclusters)
\cite{ATLAS-LARG-PUB-2008-002} that
group together cells based on their signal-to-noise ratio. The default jet
algorithm in ATLAS is the anti-$k_t$ algorithm~\cite{antikT,Cacciari:2006}.
For the jet collection used in this analysis, the distance parameter of $R = 0.6$ is chosen.
Jets are first calibrated at the electromagnetic scale (EM calibration), which accounts
correctly for the energy deposited by electromagnetic showers but does
not correct the scale for hadronic showers.

The hadronic calibration is applied in steps,
using a combination of techniques based on Monte Carlo (MC) simulation
and {\it in situ} measurements \cite{JESUncertaintyR17}.  
The first step is the pile-up correction which accounts
for the additional energy due to collisions in the same bunch crossing
as the signal event (in-time) or in nearby bunch crossings (out-of-time).  
Since the pile-up is a combination of these effects, the net correction
may add or subtract energy from the jet.
In the second step, the position of the jet origin is corrected for differences
between the geometrical centre of the detector and the collision vertex.
The third step is a jet energy correction using factors that are functions
of the jet energy and pseudorapidity. These calibration factors are derived
from MC simulation using a detailed description of the ATLAS 
detector geometry, which simulates the main detector response effects.
The EM and hadronic calibration steps above are referred to collectively
as the ``EM+JES'' scheme~\cite{JESPaper}, which restores the 
hadronic jet response in MC to within 2\%. 

The level of agreement between data and MC simulation is further improved by the
application of calibration steps based on {\it in situ} studies.
First, the relative response in $|\eta|$ is equalised using an inter-calibration
method obtained from balancing the transverse momenta of jets in dijet
events \cite{EtaIntercalibration}. 
Then the absolute energy response is brought into closer agreement with MC
simulation by a combination of various techniques based on momentum balancing
methods between photons or $Z$ bosons and jets, and between high-momentum jets and a recoil 
system of low-momentum jets.  This completes all the stages of the jet calibration.
  
The jet energy scale uncertainty is determined for jets with transverse momenta above
20~GeV and $|\eta| <$ 4.5, based on the uncertainties of the {\it in situ} techniques
and on systematic variations in MC simulations.
For the most general case, covering all jet measurements made in ATLAS,
the correlations among JES uncertainties are described by a set of 58
sources of systematic uncertainty (nuisance parameters).  Uncertainties due to pile-up, jet flavour,
and jet topology are described by five additional nuisance parameters. 
The total uncertainty from {\it in situ} techniques for central jets with a transverse momentum of
100~GeV is  as low as 1\% and  rises to about 4\% for
jets with transverse momentum above 1~TeV.

For the high-\pT\ dijet measurements made in the current analysis, the number of nuisance parameters
is reduced to 14, while keeping a correlation matrix and total magnitude 
equivalent to the full configuration. This is achieved using a procedure that 
diagonalises the total covariance matrix found from {\it in situ} techniques,
selects the largest eigenvalues as effective nuisance parameters,
and groups the remaining parameters into one additional term.

The jet energy resolution is estimated both in data and in simulation using transverse
momentum balance studies in dijet events, 
and they are found to be in good agreement~\cite{JetResConf2012}.
Monte Carlo studies are used to assess the dijet mass resolution. 
Jets constructed from final state particles are compared to the calorimeter jets
obtained after the same particles have been passed through full detector simulation. 
While the dijet mass resolution is found to be 10\% at 0.20 TeV, it is reduced to
approximately 5\% within the range of high dijet masses considered in the current studies.

\section{Event selection criteria}
\label{sec:recocuts}

The triggers employed for this study select events that have at least
one large (100~GeV or more) 
transverse energy deposition in the calorimeter.  These triggers are also referred to
as ``single jet'' triggers.
To match the data rate to the processing and storage capacity available to ATLAS, a number
of triggers with low-\pT\ thresholds were ``prescaled''.
For these triggers only a preselected fraction of all events passing
the threshold is recorded. 

A single, unprescaled trigger is used for the dijet mass spectrum analysis. This single
trigger is also used for the angular analyses at high dijet mass, but in addition several
prescaled triggers are used at lower dijet masses. Each $\chi$ distribution is assigned
a unique trigger, chosen to maximise the statistics, leading to a different effective
luminosity for each distribution. Similar choices are made for the \Fchimjj\ distribution,
assigning triggers to specific ranges of $\mjj$ to maximise the statistics in each range. 
In all analyses, kinematic selection criteria ensure a trigger efficiency
exceeding 99\% for the events under consideration.

Events are required to have a primary collision vertex defined by two or more charged
particle tracks. In the presence of additional $pp$ interactions, the primary collision
vertex chosen is the one with the largest scalar sum of $p_{\mathrm{T}}^2$ for its associated 
tracks. In this analysis, the two highest-\pT\ jets are invariably associated with this largest
sum of $p_{\mathrm{T}}^2$ collection of tracks, which ensures that the correct collision
vertex is used to reconstruct the dijet.
Events are rejected if the data from the electromagnetic calorimeter have a topology as expected
for non-collision background, or there is evidence of data corruption~\cite{JetEtMiss_Cleaning_Note}.
There must be at least two jets within $|y| < 4.4$ in the event, and
all jets with $|y| \geq 4.4$ are discarded. The highest \pT\ jet is referred to as the ``leading jet'' ($j_1$),
and the second highest as the ``next-to-leading jet'' ($j_2$).  
These two jets are collectively referred to as the ``leading jets''. 
Following the criteria in ref.~\cite{JetEtMiss_Cleaning_Note},
there must be no poorly measured jets with \pT\ greater than 30\% of
the \pT\ of the next-to-leading jet for events to be retained. 
Poorly measured jets correspond to energy depositions in regions where the energy
measurement is known to be inaccurate. Furthermore, if either of the leading jets is not
attributed to in-time energy depositions in the calorimeters, the event
is rejected.

A selection has been implemented to avoid a defect in the readout electronics of
the electromagnetic calorimeter in the
region from $-0.1$ to $1.5$ in $\eta$, and from $-0.9$ to $-0.5$ in $\phi$ that
occurred during part of the running period.  The average response for
jets in this region is 20\% to 30\% too low. 
For the $\mjj$ analysis,  events in the affected running period with jets near this
region are rejected if such jets have a \pT\ greater than
30\% of the next-to-leading jet \pT.
This requirement removes 1\% of the events. 
A similar rejection has been made for the angular analysis.  In this case
the complete $\eta$ slice from $-0.9$ to $-0.5$ in $\phi$ is excluded in
order to retain the shape of the distributions.
The event reduction during run periods affected
by the defect is 13\%, and the overall reduction in the data set due to
this effect is 4\%.  

Additional kinematic selection criteria are used to enrich the sample with events
in the hard-scattering region of phase space.  
For the dijet mass analysis, events must satisfy 
$|y^*| < 0.6$\, and $|\eta_{1,2}| < 2.8$ for the leading jets, and
$\mjj > 850$~GeV.

For the angular analyses, events must satisfy 
$|y^*| < 1.7$\, and $|y_B| < 1.1$, and $\mjj > 800$~GeV. 
The combined $y^*$ and $y_B$ criteria  limit the rapidity range of the leading jets to 
$|y_{1,2}| < 2.8$.
This  $|y_B|$ selection does not affect events with dijet mass
above 2.8~TeV since the phase space is kinematically constrained. 
The kinematic selection also restricts the minimum \pT\ of jets entering the analysis 
to 80~GeV. 
Since at lowest order $y_B = \frac{1}{2} \ln(\frac{x_1}{x_2})$ and
$\mjj^2 = x_1\, x_2 \, s$, with $x_{1,2}$ the parton momentum fractions
of the colliding protons, the combined $\mjj$ and $y_B$
criteria result in limiting the effective $x_{1,2}$-ranges in the
convolution of the matrix elements with the PDFs. 
The QCD matrix elements for dijet production lead to $\chi$ distributions that are approximately flat.
Without the selection on $y_B$, the $\chi$ distributions predicted by QCD would have
a slope becoming more pronounced for the lower $\mjj$ bins. 
Restricting the $x_{1,2}$-ranges of the PDFs reduces this shape distortion, and also
reduces the PDF and jet energy scale uncertainties associated with each $\chi$ bin
of the final distribution.

\section {Comparing the dijet mass spectrum to a smooth background}
\label{sec:smoothfit}

\begin{figure}[!htb]
  \centering \includegraphics[width=0.70\textwidth]{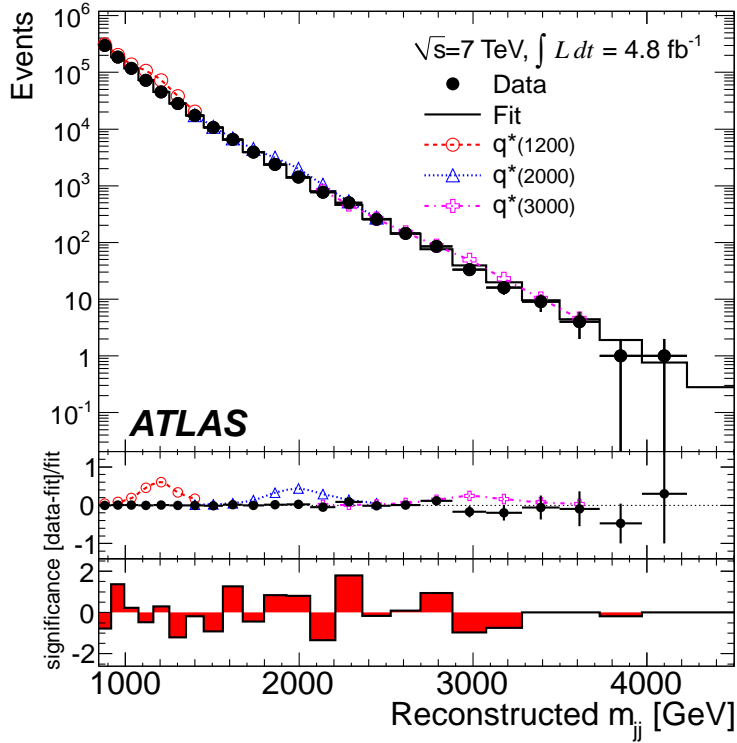}
  \caption{The reconstructed dijet mass distribution (filled points) fitted with
  a smooth functional form (solid line).
  Mass distribution predictions for three $\qstar$ masses are shown above the background.
  The middle part of figure shows the data minus the background fit, divided by the fit.
  The bin-by-bin significance of the data-background difference is shown in the lower panel.}
  \label{fig:massdist}
\end{figure}

In the dijet mass analysis, a search for resonances in the $\mjj$
spectrum is made by using a data-driven background estimate.
The observed dijet mass distribution after all selection cuts is shown
in figure \ref{fig:massdist}. Also shown in the figure are the predictions
for an excited quark for three different mass hypotheses \cite{Baur:1987ga,Baur:1989kv}.  
The $\mjj$ spectrum is fit to a smooth functional form,
\begin{equation}
  f(x) = p_1 (1 - x)^{p_2} x^{p_3 + p_4\ln x},
\label{Eq:fitfunction}
\end{equation}
where the $p_i$\ are fit parameters, and $x \equiv \mjj/\sqrt{s}$. 
In previous studies, ATLAS and other
experiments~\cite{CDF:2009DijetSearch,ATLAS:2010bc,CMS:2010dijetmass,Aad:2011aj}
have found this ansatz to provide a satisfactory
fit to the QCD prediction of dijet production.
The use of a full Monte Carlo QCD background prediction would 
introduce theoretical and systematic uncertainties of its own,
whereas this smooth background form introduces only the
uncertainties associated with its fit parameters.
A feature of the functional form used in the fitting is that it allows for
smooth background variations but does not accommodate
localised excesses that could indicate the presence of NP signals.
However, the effects of smooth deviations from QCD, such as contact
interactions, could be absorbed by the background fitting
function, and therefore the  $\mjj$ analysis is used only to search
for resonant effects.

\interfootnotelinepenalty=1000

The $\chi^2$-value of the fit is 17.7 for 22 degrees of freedom, and the reduced $\chi^2$ is 0.80. The middle part of figure \ref{fig:massdist} shows the data minus the
background fit, divided by the fit.
The lower part of figure \ref{fig:massdist} shows the significance, in standard deviations, 
of the difference between the data and the fit in each bin.
The significance is calculated taking only statistical uncertainties into account, and assuming that the data follow a Poisson distribution.
For each bin a p-value is determined by assessing the probability of the background 
fluctuating higher than the observed excess or lower than the observed
deficit.  This $p$-value is transformed to a significance in terms of an equivalent number
of standard deviations (the $z$-value)~\cite{PlotDiffs}.
Where there is an excess (deficit) in data in a given bin, the significance is plotted as
positive (negative)\footnote{
In mass bins with small expected number of events, where the observed number of events is similar
to the expectation, the Poisson probability of a fluctuation at least as high (low) as the observed
excess (deficit) can be greater than 50\%, as a result of the asymmetry of the Poisson distribution.
Since these bins have too few events for the significance to be meaningful, the bars are not drawn for them.}.
To test the degree of consistency between the data and the fitted background, the $p$-value
of the fit is determined by calculating the $\chi^2$-value from the data and comparing this result to the
$\chi^2$ distribution obtained from pseudo-experiments drawn from
the background fit, as described in a previous
publication~\cite{Aad:2011aj}.  The resulting $p$-value is 0.73, 
showing that there is good agreement between the data and the fit.

As a more sensitive test, the \BumpHunter\ algorithm~\cite{Aaltonen:2008vt,Choudalakis:2011bh}
is used to establish the presence or absence of a resonance in the dijet mass spectrum,
as described in greater detail in previous publications~\cite{Aad:2011aj,ATLAS:Res2011}.
Starting with a two-bin window, the algorithm
increases the signal window and shifts its location until all possible bin ranges,
up to half the mass range spanned by the data, have been tested.  
The most significant departure from the smooth spectrum (``bump'') is defined by the
set of bins that have the smallest probability of arising from a background
fluctuation assuming Poisson statistics.

The \BumpHunter\ algorithm accounts for the so-called ``look-elsewhere effect'' 
~\cite{lookelsefirstcite}, by performing a
series of pseudo-experiments drawn from the background estimate to determine the probability that random
fluctuations in the background-only hypothesis would create an excess 
anywhere in the spectrum at least as significant as the one observed.
Furthermore, to prevent any NP signal from biasing the background estimate,
if the most significant local excess from the background fit has a $p$-value smaller than 0.01,
this region is excluded and a new background fit is performed.  No such exclusion
is needed for this data set.

The most significant discrepancy identified by the \BumpHunter\ algorithm in the
observed dijet mass distribution in figure \ref{fig:massdist} is a four-bin excess in
the interval 2.21~TeV to 2.88~TeV.  The probability of observing such an excess or larger
somewhere in the mass spectrum for a background-only hypothesis is 0.69. 
This test shows no evidence for a resonance signal in the $\mjj$\ spectrum.

\section {QCD predictions for dijet angular distributions}
\label{Sec:QCDPredic}

In the dijet angular analyses, the QCD prediction is based on MC generation of
event samples which cover the kinematic range in $\chi$ and $\mjj$ spanned by
the selected dijet events. The QCD hard scattering interactions are simulated using
the \Pythia~6 \cite{Pythia6} 
event generator with the ATLAS AUET2B LO** tune~\cite{ATLAS_MC11bc} which uses the
MRSTMCal~\cite{Sherstnev:2007nd} modified leading-order (LO) parton distribution functions (PDFs).

To incorporate detector effects, these QCD events are passed through a fast detector simulation,
ATLFAST~2.0~\cite{ATLFAST2-1998}, which employs FastCaloSim~\cite{ATLFAST2-2011} for the
simulation of electromagnetic and hadronic showers in the calorimeter.
Comparisons with detailed simulations of the ATLAS
detector~\cite{ATLSIM,Agostinelli:2002hh} using the \Geant\
package~\cite{Agostinelli:2002hh} show no differences in the angular
distributions exceeding 5\%. 

To simulate in-time pile-up, separate samples of inelastic interactions are
generated using \Pythia~8~\cite{Pythia8}, and these samples
are passed through the full detector simulation.  
To simulate QCD events in the presence of pile-up, hard scattering events are overlaid
with $\mu$ inelastic interactions, where $\mu$ is Poisson distributed, and
the distribution of $\langle\mu \rangle $ is chosen to match the distribution of average number
of interactions per bunch crossings in data. 
The combined MC events, containing one hard interaction and several soft interactions,
are then reconstructed in the same way as collision data and are subjected to the same event selection
criteria as applied to collision data.

Bin-by-bin correction factors (K-factors) are applied
to the angular distributions derived from MC calculations to account
for NLO contributions. 
These K-factors are derived from dedicated MC samples and are defined as
the ratio $NLO_{ME}$/$PYT_{SHOW}$. 
The $NLO_{ME}$ sample is produced using NLO matrix elements in \NLOJET\
\cite{Nagy1,Nagy2,catani-1998-510} 
with the NLO PDF from CT10~\cite{CTEQ10}.
The $PYT_{SHOW}$ sample is produced with the \Pythia~6 generator restricted to 
leading-order matrix elements and with parton showering but with
non-perturbative effects turned off. This sample also uses the AUET2B LO** tune.

The angular distributions generated with the full \Pythia\ simulation include
various non-perturbative effects including hadronisation, underlying event, 
and primordial $k_{\perp}$.
The K-factors defined above are designed to retain these effects while
adjusting for differences in the treatment of perturbative effects.
The full \Pythia\ predictions of angular distributions are multiplied by these
bin-wise K-factors to obtain  reshaped spectra that include corrections
originating from NLO matrix elements.  
K-factors are applied to $\chi$ distributions before normalising them
to unit area.
The K-factors change the normalised 
$\chi$ distributions by 2\% at low dijet mass, by as much as 11\%
in the highest dijet mass bins, and the
effect is largest at low $\chi$. 
The K-factors for  \Fchimjj\ are close to unity for dijet masses of
around 1~TeV, but increase with dijet mass, and are as large as 20\% for dijet masses of 4~TeV. 
Electroweak corrections are not included in the theoretical predictions \cite{ewcorrs}.

\section {Comparing $\chi$ distributions to QCD predictions}
\label{sec:compare:chi}

\begin{figure}[h!]
  \centering \includegraphics[width=0.95\textwidth]{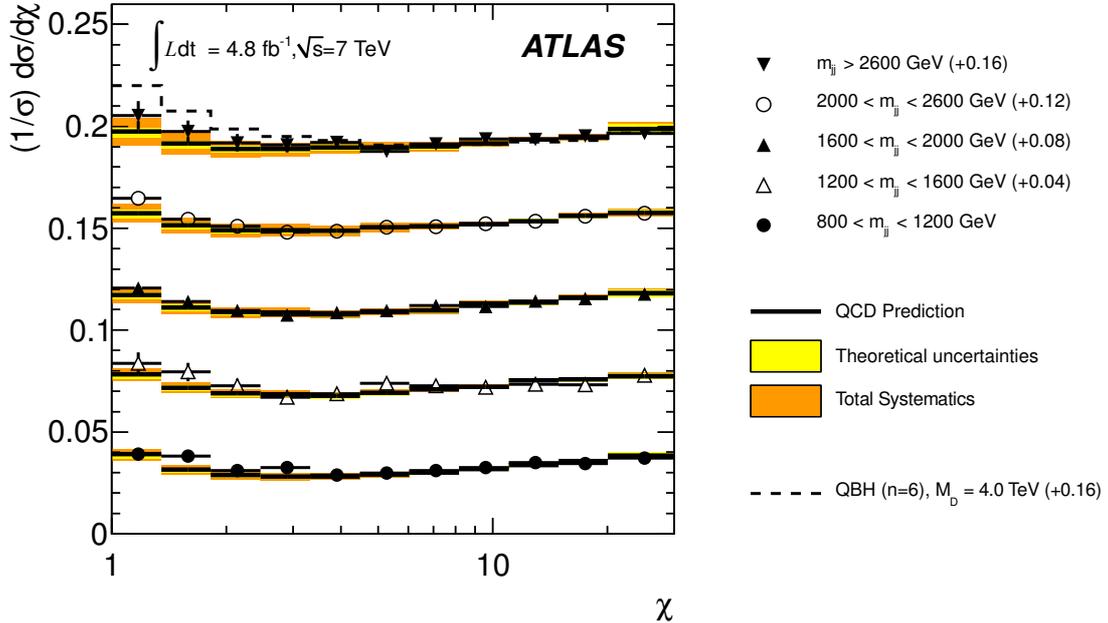}
  \caption{The $\chi$ distributions for all dijet mass bins. 
      The QCD predictions are shown with theoretical and total systematic uncertainties (bands),
      as well as the data with statistical uncertainties. 
      The dashed line is the prediction for
      a QBH signal for $M_D =4.0$~TeV and $n = 6$\ in the highest mass bin.
      The distributions have been offset by the amount shown in the 
      legend to aid in visually comparing the shapes in each mass bin. 
      }
  \label{fig:chisvsQCD}
\end{figure}

The observed $\chi$\ distributions normalised to unit area are shown
in figure \ref{fig:chisvsQCD}\ for several
$\mjj$\ bins, defined by boundaries at 800, 1200, 1600, 2000, and
2600~GeV.
The highest bin includes all dijet events with $\mjj > 2.6$~TeV.
The dijet mass bins are chosen to ensure sufficient entries in each mass
bin. From the lowest dijet mass bin to the highest bin, the number of
events are: 13642,  4132, 35250, 28462, 2706, and the corresponding
integrated luminosities are \mbox{5.6 pb$^{-1}$}, \mbox{19.2 pb$^{-1}$}, \mbox{1.2
fb$^{-1}$}, \mbox{4.8 fb$^{-1}$} and \mbox{4.8 fb$^{-1}$}. 
The yield for all \mbox{$\mjj < 2000$~GeV} is reduced due to the usage of prescaled triggers, 
and for \mbox{$\mjj > 2000$~GeV} by the falling cross section.

The $\chi$\ distributions are compared to the predictions from QCD,
which include all systematic uncertainties, 
and the signal predictions of one particular NP model,
a quantum black hole (QBH) scenario
with a quantum gravity mass scale of 4.0~TeV and six extra dimensions \cite{RandallMeade,Feng:2004}.

Pseudo-experiments are used to convolve statistical, systematic and
theoretical uncertainties on the QCD predictions, as has been done
in previous studies of this type~\cite{ATLAS:2010eza}.  The primary sources of theoretical 
uncertainty are NLO QCD renormalisation and factorisation scales, 
and PDF uncertainties. The QCD scales are varied by a factor of two
independently around their nominal values, which are set to the mean
\pT\ of the leading jets, while the
PDF uncertainties are determined using CT10 NLO PDF error sets~\cite{LHCprimer}.
The resulting bin-wise uncertainties for the cross-section normalised $\chi$ distributions
can be as high as 8\% for the combined NLO QCD scale variations and are typically 
below 1\% for the PDF uncertainties. These theoretical uncertainties are convolved
with the JES uncertainty and applied to all MC angular distributions. 
Other experimental uncertainties such as those due to pile-up and
to the jet energy and angular resolutions have been investigated and found
to be negligible. 
The JES uncertainties are largest at low $\chi$ and are as small as 5\% for the lowest dijet mass
bin but increase to above 15\% for the highest bin.
Variations based on the resulting systematic uncertainties are used in generating statistical ensembles
for the estimation of $p$-values when comparing QCD predictions to data.

A statistical analysis is performed on each of the five $\chi$ distributions to test
the overall consistency between data and QCD predictions.
A binned log-likelihood is calculated for each distribution assuming
that the sample consists only of QCD dijet production. The expected distribution of this
likelihood is then determined using pseudo-experiments drawn from the QCD MC sample and
convolved with the systematic uncertainties as discussed above.
Finally the $p$-value is defined as the probability of obtaining a log-likelihood value
less than the value observed in data.   

The $p$-values determined from the observed likelihoods are shown in
table \ref{tab:chitoqcd}. 
These indicate that there is no statistically significant evidence for new
phenomena in the $\chi$ distributions, and that these distributions are
in reasonable agreement with QCD predictions.  

As with the dijet resonance analysis, the \BumpHunter\ algorithm is applied to the five $\chi$ distributions separately, in
this case to test for the presence of features that might indicate
disagreement with the QCD prediction. The results are shown in table \ref{tab:chitoqcd}.
In this particular application, the  \BumpHunter\ is required to start
from the first $\chi$ bin, and the excess must be at least three bins
wide. For each of the bin combinations, the binomial $p$-value for
observing the data given the QCD-background-only hypothesis is calculated. 
The bin sequence with the smallest binomial $p$-value is listed in
table \ref{tab:chitoqcd}.  
Statistical and systematic uncertainties, and look-elsewhere effects, are included using
pseudo-experiments drawn from the QCD background. For each of the
pseudo-experiments the most discrepant bin combination is found and
its $p$-value is used to construct the expected binomial $p$-value distribution. The final  \BumpHunter\  $p$-value is then defined as the probability of finding a binomial $p$-value as extreme as the one observed in data. 
The $p$-values listed in the last column of table \ref{tab:chitoqcd}
indicate that the data are consistent with the
QCD prediction in all five mass bins.

\begin{table}[h!]
\centering {
\begin{tabular}{|l|l|ll|}
\hline
$\mjj$ bin    &  LL       &      BH    &          BH               \\ 
  $[\mathrm{GeV}]$       &   $p$-value   &   Discrep  &    $p$-value \\ \hline
    800--1200     &   0.23    &   bin 1--9  &      0.17 \\
   1200--1600     &   0.31    &   bin 1--7  &      0.20  \\
   1600--2000     &   0.56    &   bin 1--7  &      0.37 \\
   2000--2600     &   0.74    &   bin 1--3  &      0.38  \\
   $>$ 2600       &   0.83    &   bin 1--10  &      0.37 \\
\hline
\end{tabular}
}
\caption {Comparing $\chi$ distributions to QCD predictions. The abbreviations in the first line of the table stand for ``log-likelihood'' (LL), and ``\BumpHunter'' (BH).
The second line labels the ``$p$-values'' ($p$-value) and the ``most discrepant region'' (Discrep).}
      \label{tab:chitoqcd}
\end{table}

In addition, the \BumpHunter\ algorithm is applied to all $\chi$ distributions
at once, which increases the effect of the correction for the look-elsewhere effect.  The most
discrepant region in all distributions is in bins 1--9 of the 800--1200~GeV mass distribution.
The resulting $p$-value, including the look-elsewhere effect, is now 0.43, 
again indicating good agreement with QCD predictions.

\section {Comparing the \Fchimjj\ distribution to the QCD prediction}
\label{sec:Fchimjj}

\begin{figure}[ht!]
  \centering \includegraphics[width=0.7\textwidth]{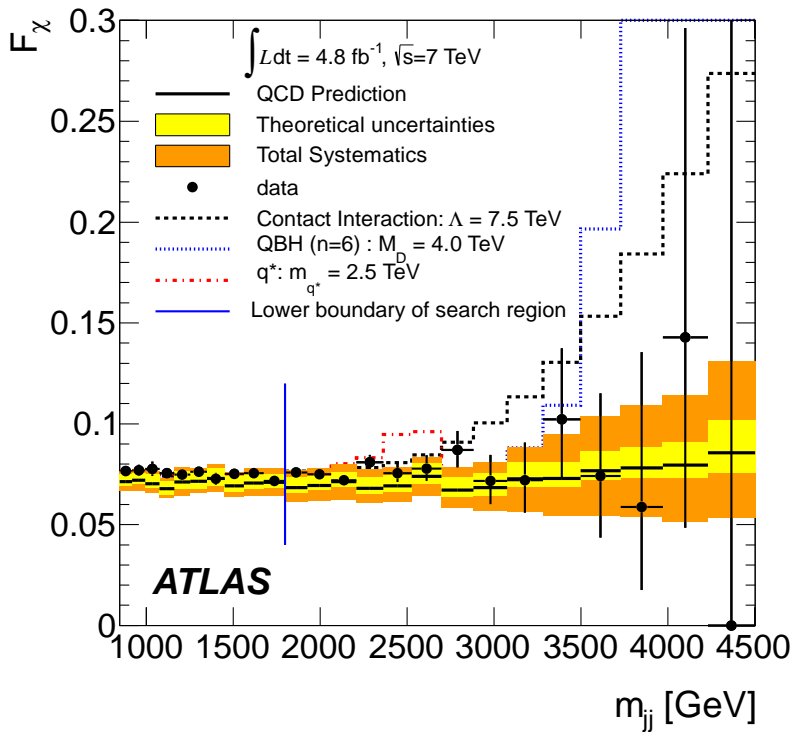}
  \caption{The \Fchimjj\ distribution in $\mjj$.  The QCD prediction is shown 
 with theoretical and total systematic uncertainties (bands), 
 and data (black points) with statistical uncertainties. The blue
 vertical line indicates the lower boundary of the search region for new phenomena.
 Various expected new physics signals are shown: a contact interaction
 with  $\Lambda$ = 7.5~TeV, an excited quark with mass 2.5~TeV and a
 QBH signal with $M_D =4.0$~TeV. 
  }
  \label{fig:fchicompQCD}
\end{figure}

The observed \Fchimjj\ data distribution is shown in figure \ref{fig:fchicompQCD}, where it
is compared to the QCD prediction, which includes all systematic uncertainties.
Also shown in the figure is the expected behaviour of \Fchimjj\ if a contact interaction
with the compositeness scale $\Lambda=7.5$~TeV were
present \cite{Eichten:1984eu,Eichten:1995akc,Chiappetta1991}. Furthermore the predictions for an excited quark with
a mass of 2.5~TeV and a QBH signal with $M_D =4.0$~TeV are shown.
The blue vertical line at 1.8~TeV included in figure \ref{fig:fchicompQCD} indicates the mass
boundary above which the search phase of the analysis is performed, as
explained below.

The observed \Fchimjj\ distribution is obtained by forming the
finely-binned $\mjj$ distributions for $N_\mathrm{central}$ and
$N_\mathrm{total}$ --- the
``numerator'' and ``denominator'' distributions of \Fchimjj\ ---
separately and taking the ratio.  The handling of systematic
uncertainties, including JES, PDF and scale uncertainties, uses a procedure similar to that for the $\chi$ distributions.

Two statistical tests are applied to the high-mass region to determine whether the data are 
compatible with the QCD prediction.
The first test uses a binned likelihood, which includes the systematic uncertainties, and is constructed assuming the presence of QCD processes only. The $p$-value calculated from this likelihood is 0.38, indicating that these data are in agreement with the QCD prediction.

The second test consists of applying the \BumpHunter\ and \TailHunter\ algorithms
~\cite{Aaltonen:2008vt,Choudalakis:2011bh} to the \Fchimjj\
distributions, including systematic uncertainties and assuming binomial statistics. 
For this test only data with dijet masses above 1.8 TeV, associated with the single unprescaled
trigger, are used to obtain a high sensitivity at high mass and to avoid diluting the
test with the large number of low-mass bins.
The test scans the data using windows of varying widths and identifies
the window with the largest excess of events with respect to the background. 
The \BumpHunter\ finds the most discrepant interval to be from 1.80~TeV
to 2.88~TeV, with a $p$-value of 0.20. 
The \TailHunter\ finds the most discrepant interval to be from 1.80~TeV
onwards, with a $p$-value of 0.21. 
The $p$-values indicate that there is no significant excess in the data .

\section {Simulation of hypothetical new phenomena}
\label{sec:SimulNP}
In the absence of any significant signals indicating the presence of phenomena
beyond QCD, Bayesian 95\% credibility level (CL) limits are determined for a number of NP hypotheses.
The following models have been described in detail in previous ATLAS dijet studies
~\cite{ATLAS:2010bc,ATLAS:2010eza,Aad:2011aj,ATLAS:Res2011}: 
  quark contact interactions (CI) \cite{Eichten:1984eu,Eichten:1995akc,Chiappetta1991},
  excited quarks ($\qstar$) \cite{Baur:1987ga,Baur:1989kv},
  colour octet scalars (s8) \cite{Han:2010rf}, and
  quantum black holes (QBH) \cite{RandallMeade,Feng:2004}. 
Two models of new phenomena are added to the current analysis:
  heavy $W$ bosons ($\Wprime$) with SM couplings~\cite{Wprime1,Wprime2,Wprime3,Wprime4},
  and string resonances (SR)~\cite{StrRes1,StrRes2,StrRes3,StrRes4}.
Contact interactions and QBH appear as slowly rising effects in $\mjj$, while
the other hypotheses produce localised excesses.  

A number of these NP models are available in the \Pythia~6 event generator.
In these cases, the corresponding MC samples are generated using
the AUET2B LO** tune and the MRSTMCal PDF.
For NP models provided by other event generators, with other PDFs, partons
originating from the initial two-parton interaction are used as input to \Pythia\,
which performs parton showering and the remaining event generation steps.
In all cases, the renormalisation and factorisation scales are set to the mean \pT\
of the leading jets. 

The quark contact interaction, CI,
is used to model the appearance of kinematic properties that characterise quark compositeness.
In the current analysis, only destructive interference is
studied, but constructive interference is expected to give less conservative limits. 
\Pythia~6 is used to create MC event samples for distinct values of the 
compositeness scale, $\Lambda$.

Excited quarks, $\qstar$ , a possible manifestation of
quark compositeness, are also simulated
in all decay modes with \Pythia~6 for selected values of the $\qstar$ mass. 
Excited quarks are assumed to decay to common quarks via standard model couplings,
leading to gluon emission approximately 83\% of the time.
Recent studies comparing this benchmark model to the same excited quark model
in \Pythia~8 show that the $\qstar$ $\mjj$ distribution in \Pythia~8 is
significantly broader than that in \Pythia~6. 
The \Pythia\ authors have identified a long-standing misapplication of QCD
\pT-ordered final state radiation (FSR) vetoing in \Pythia~6, which is resolved
in \Pythia~8.  
The $\qstar$ $\mjj$ distributions from \Pythia~6 can be brought into close 
correspondence with \Pythia~8 by setting the \Pythia~6 MSTJ(47) 
parameter to zero, restoring the correct behaviour for final state radiation.
The resulting widening of the peak affects the search sensitivity and exclusion limits.
The $\qstar$ MC samples used in the current studies are generated 
using both the default and corrected \Pythia~6 settings, to determine the impact
on the $\qstar$ exclusion limit.

The colour octet scalar model, s8, is a typical example of 
possible exotic coloured resonances 
decaying to two gluons.
{\sc MadGraph}~5~\cite{Alwall:2011uj} with the CTEQ6L1 PDF \cite{Pumplin:2002vw}
is employed to generate parton-level event samples at leading-order approximation
for a selection of s8 masses, which are used as input to \Pythia~6.

A model for quantum black holes, QBH,  
that decay to two jets is simulated using \BlackMax~\cite{Dai:2007bm}\ with the CT10 PDF
to produce a simple two-body final state scenario of quantum gravitational effects at the
reduced Planck Scale $M_D$, with $n$ = 6 extra spatial dimensions.
The QBH model is used as a benchmark to represent any quantum gravitational effect that
produces events containing dijets.  Event samples for selected values of $M_D$ are
used as input to \Pythia\ for further processing.

The first new NP phenomenon used in the current dijet analysis, 
the production of heavy charged gauge bosons, $\Wprime$, has been sought
in events containing a charged lepton (electron or muon) and a 
neutrino~\cite{Wprime3,Wprime4}, but no evidence has been found. In the current studies,
dijet events are searched for the decays of $\Wprime$ to $q\bar{q}'$.  The specific model
used in this study~\cite{Wprime1,Wprime2} assumes that the $\Wprime$
has V-A SM couplings but does not include interference between the $\Wprime$ and the $W$.
The $\Wprime$ signal sample is generated with the \Pythia~6 event generator.
Instead of the LO cross section values, the NNLO electroweak-corrected cross section values
\cite{Wprime4,Hamberg:1990np, CarloniCalame:2006zq,CarloniCalame:2007cd} calculated using
the MSTW2008 PDF \cite{Martin:2009iq}, are used in this analysis.
For a given $\Wprime$ mass, the width of the resonance in $\mjj$ is very similar to that
of the $\qstar$, and the angular distribution peaks at low $\chi$.  
The limit analysis for this $\Wprime$ model includes the branching ratio to the chosen
$q\bar{q}'$ final state and, for each simulated mass, this fraction is taken from
\Pythia~6.

The second new NP model considered, string resonances (SR), results from excitations of 
quarks and gluons at the string level~\cite{StrRes1,StrRes2,StrRes3,StrRes4}. 
The dominant decay mode is to $qg$, and the SR model described in ref.~\cite{StrRes3}
is implemented in the \CalcHEP\ generator \cite{Pukhov:2004ca} with the MRSTMCal PDF.
As with other models, MC samples are created for selected values of the mass parameter,
$m_{\textrm{SR}}$, by passing the \CalcHEP\ output at parton level to \Pythia~6.

All MC signal samples are passed through fast detector
simulation using ATLFAST~2.0, except for string resonances, which
are fully simulated using \Geant.

\section{Limits on new resonant phenomena from the $\mjj$ distribution}
\label{sec:mjj:reson}
For each NP process under study, Monte Carlo samples have been
simulated at a number of selected mass points, $m_{\textrm{NP}}$.
The Bayesian method documented in ref.~\cite{Aad:2011aj}
is applied to data at these same mass points to set a
95\% CL limit on the cross section times acceptance,
$\sigma\times {\cal A}$, for the NP signal as a function of
$m_{\textrm{NP}}$, using a prior constant in signal strength.
The limit on $\sigma\times {\cal A}$ from data is interpolated
between mass points to create a continuous curve in $\mjj$.
The exclusion limit on the mass (or energy scale) of the
given NP signal occurs at the value of $\mjj$ where the limit
on $\sigma\times {\cal A}$ from data is the same as the
theoretical value, which is derived by interpolation between
the generated mass values.

This form of analysis is applicable to all resonant phenomena
where the NP couplings are strong compared to the scale of perturbative QCD
at the signal mass, so that interference with QCD terms can be neglected.
The acceptance calculation includes all reconstruction steps and analysis cuts
described in section~\ref{sec:recocuts}. 
For all resonant models except for the $\Wprime$, all decay modes have been
simulated so that the branching ratio into dijets is implicitly included in
the acceptance through the analysis selection. For the $\Wprime$
model, only dijet final states have been simulated, and the branching
ratio is included in the cross section instead of in the acceptance.

\begin{figure*}[!htb]
  \centering
    \subfigure[Excited-quark model.]{
      \label{fig:limqstarmjj}
      \includegraphics[width=0.48\textwidth]
        {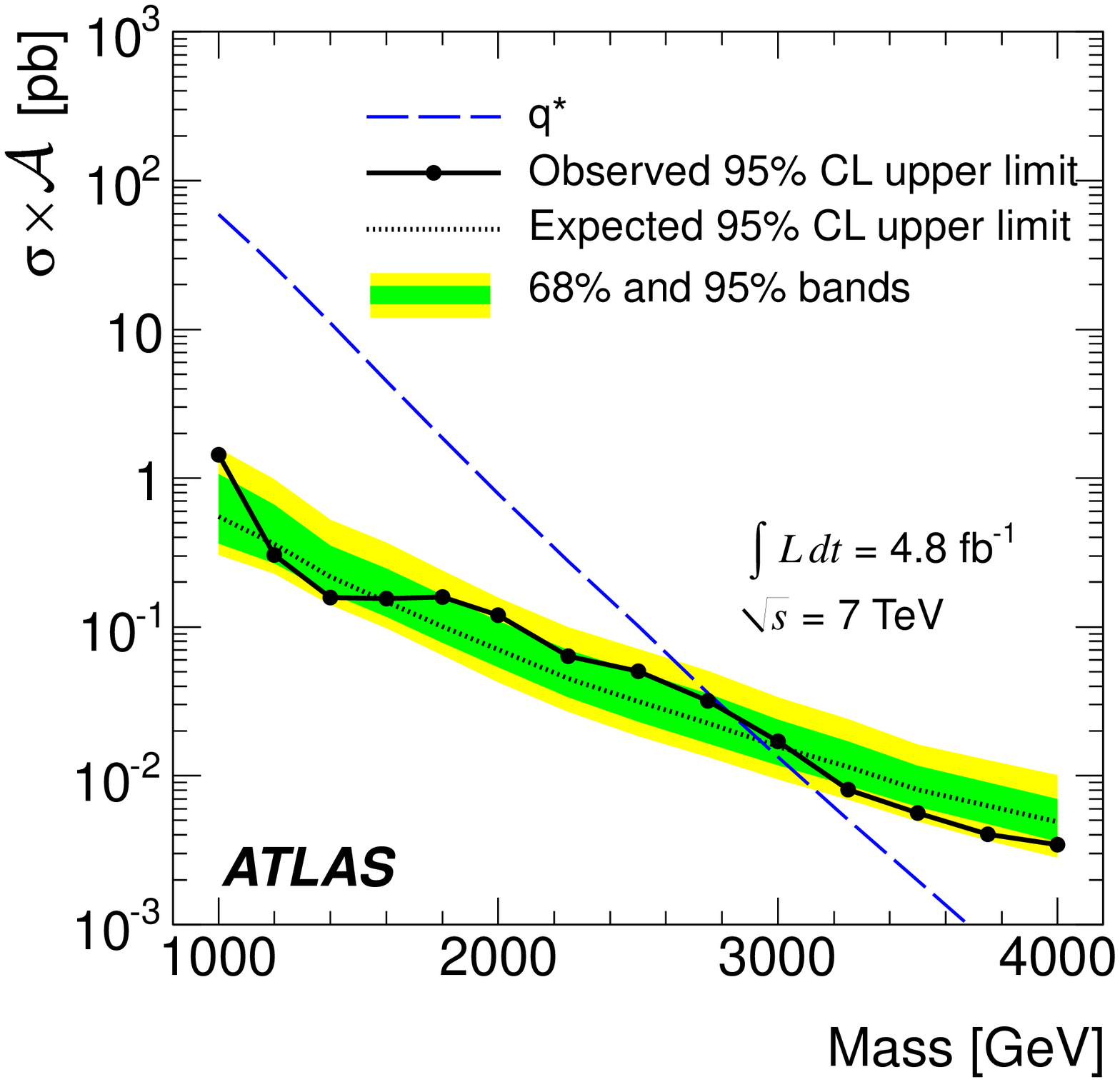}}
    \subfigure[Colour scalar octet model.]{
      \label{fig:lims8mjj}
      \includegraphics[width=0.48\textwidth]
        {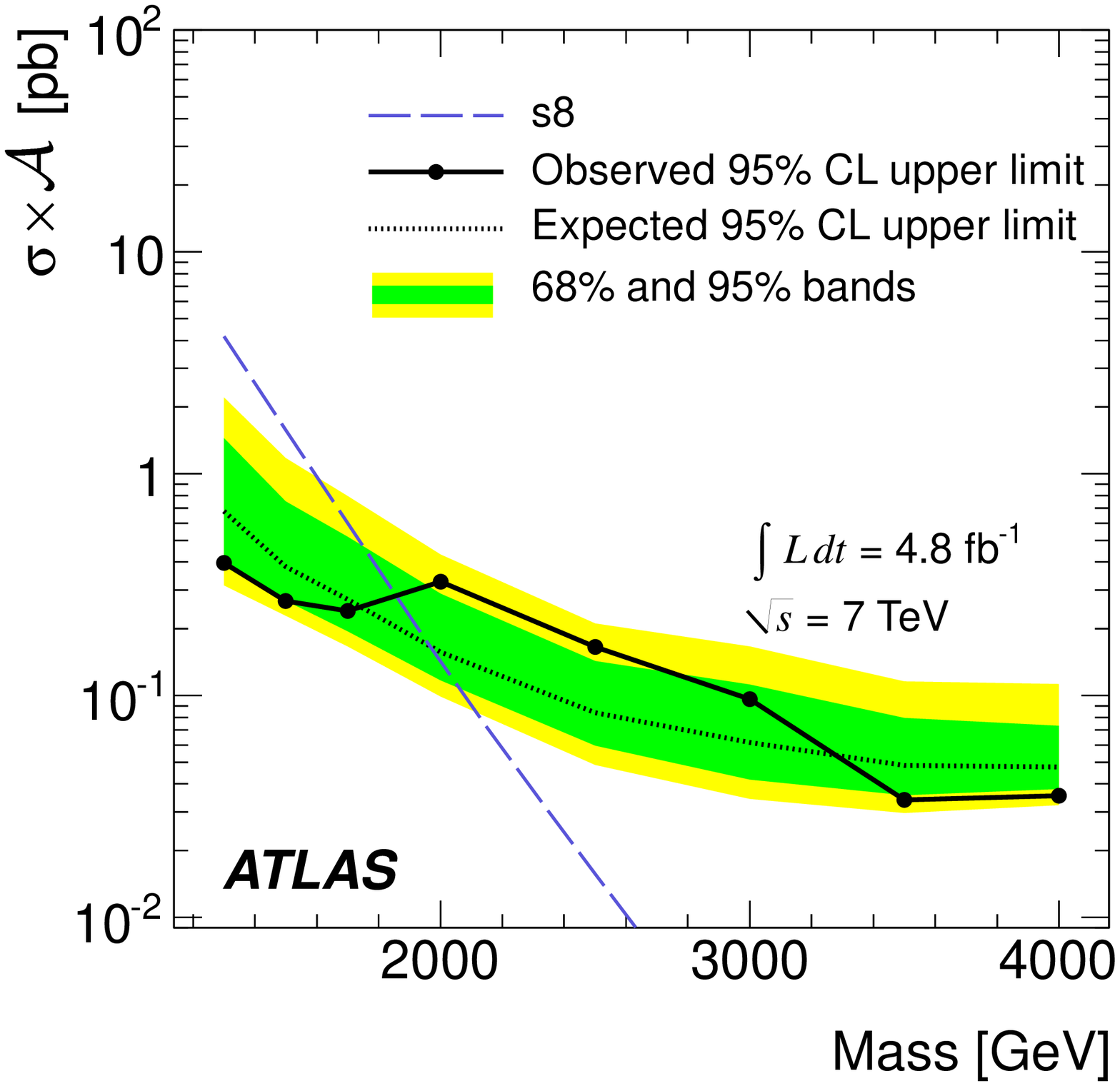}}
    \caption{The 95\% CL upper limits on $\sigma\times {\cal A}$ as a function of 
                particle mass (black filled circles)  using $\mjj$. The black dotted curve
                shows the 95\% CL upper limit expected in the absence of any
                resonance signal, and the 
		green and yellow bands represent the 68\% and 95\% contours
		of the expected limit, respectively. 
		Theoretical predictions of $\sigma\times {\cal A}$ are
                shown (dashed) in 
		(a) for excited quarks, and in (b) for colour octet scalars.
		For a given NP model, the observed (expected) limit occurs at
		the crossing of the dashed $\sigma\times {\cal A}$ curve with the observed
		(expected) 95\% CL upper limit curve.}
  \label{fig:combinedfig}
\end{figure*}

The effects of systematic uncertainties due to luminosity, acceptance, and jet energy
scale are included.
The luminosity uncertainty for the 2011 data is 3.9\%~\cite{ATLAS-CONF-2011-116} and is
combined in quadrature with the acceptance uncertainty.
The correlated systematic uncertainties corresponding to the 14 JES nuisance parameters are
added in quadrature and represented by a single nuisance parameter
which shifts the resonance mass peaks by less than 4\%.
The background parameterisation uncertainty is taken from the fit results,
as described in ref.~\cite{Aad:2011aj}.
The effect of the jet energy resolution uncertainty is found to be negligible.  

These uncertainties are incorporated into the fit by varying all sources according
to Gaussian probability distributions and convolving them with the 
posterior probability distribution. Credibility intervals are then calculated
numerically from the resulting convolutions.  No uncertainties are associated with
the theoretical model, as in each case the NP model is a benchmark that
incorporates a specific choice of model parameters, PDF set, and MC tune.  
Previous ATLAS studies using the $\qstar$ theoretical
prediction~\cite{Aad:2011aj} showed that the variation among three
different choices of MC tune and PDF set was less than 4\% for the
expected limits. 

The resulting limits for excited quarks, based on the corrected \Pythia~6
samples (as explained in section \ref{sec:SimulNP}), are shown in
figure \ref{fig:limqstarmjj}. 
The acceptance $\cal A$ ranges from 40\% to 51\% for $m_\qstar$
between 1.2~TeV and 4.0~TeV, and is never lower than 46\% for masses above 1.4~\rm TeV. 
The largest reduction in acceptance arises from the rapidity selection criteria.
The expected lower mass limit at 95\% CL for $\qstar$ is
\MjjLimitExpectedqstar~TeV, and the observed limit is \MjjLimitqstar~TeV.
For comparison, this limit has also been determined using \Pythia~6
samples with the default $\qstar$ settings, leading to narrower mass peaks.
The expected limit determined from these MC samples is 0.1~TeV higher than
the limit based on the corrected samples.
This shift is an approximate indicator of the fractional correction
that is expected when comparing the current ATLAS results to all
previous analyses that found $\qstar$ mass limits
using \Pythia~6 and \pT-ordered final state radiation without
corrections, including all previous ATLAS results. 

The limits for colour octet scalars are shown in
figure \ref{fig:lims8mjj}. 
The expected mass limit at 95\% CL is
\MjjLimitExpecteds8~TeV, and the observed limit is \MjjLimits8~TeV. 
For this model the acceptance values vary between 34\% and 48\% for masses
between 1.3~TeV and 4.0~TeV. 

\begin{figure*}[!htb]
  \centering
    \subfigure[Heavy charged gauge bosons, $\Wprime$.]{
      \label{fig:limwprimemjj}
      \includegraphics[width=0.48\textwidth]
        {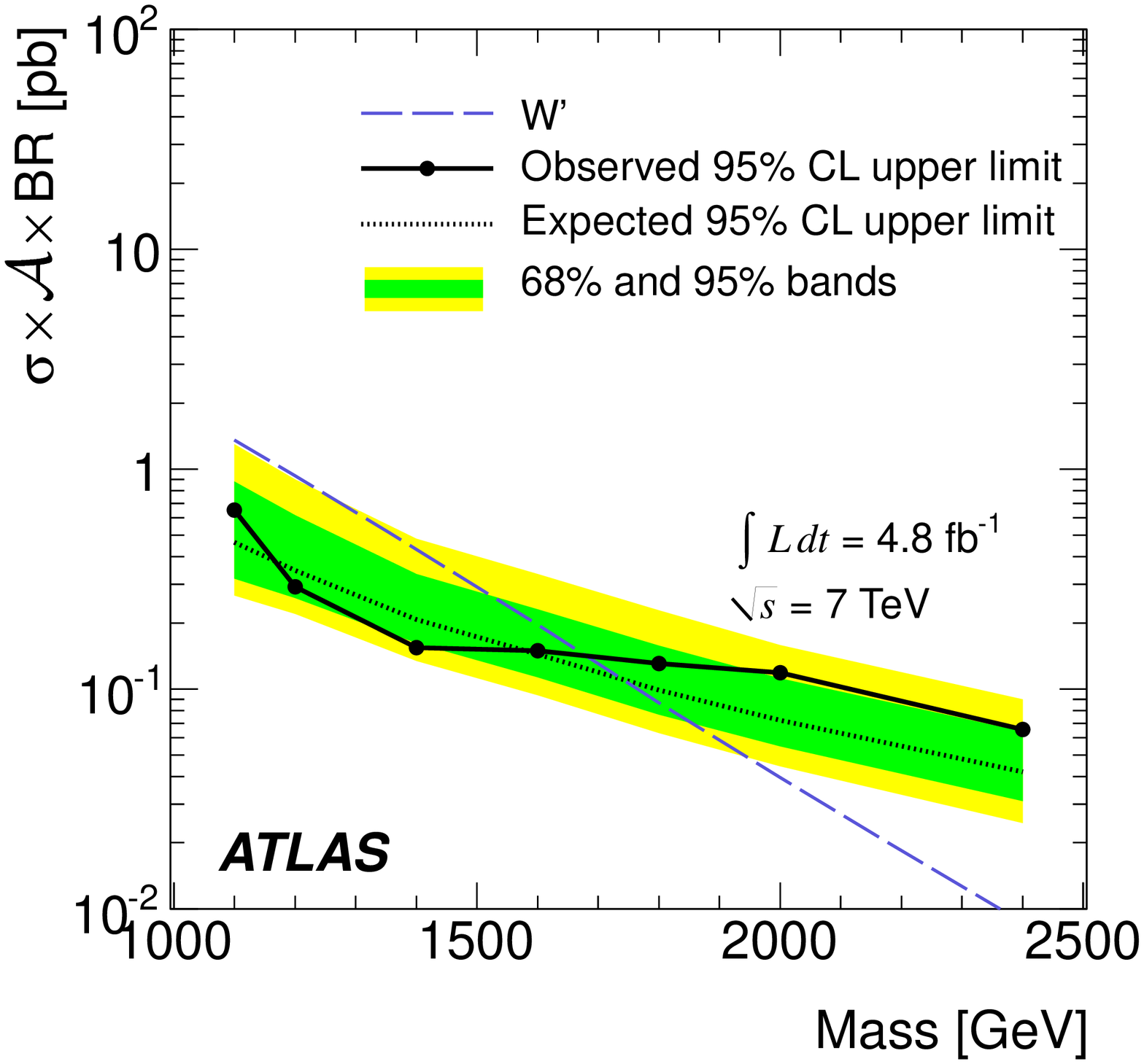}}
    \subfigure[String resonances, SR.]{
      \label{fig:limsrmjj}
      \includegraphics[width=0.48\textwidth]
        {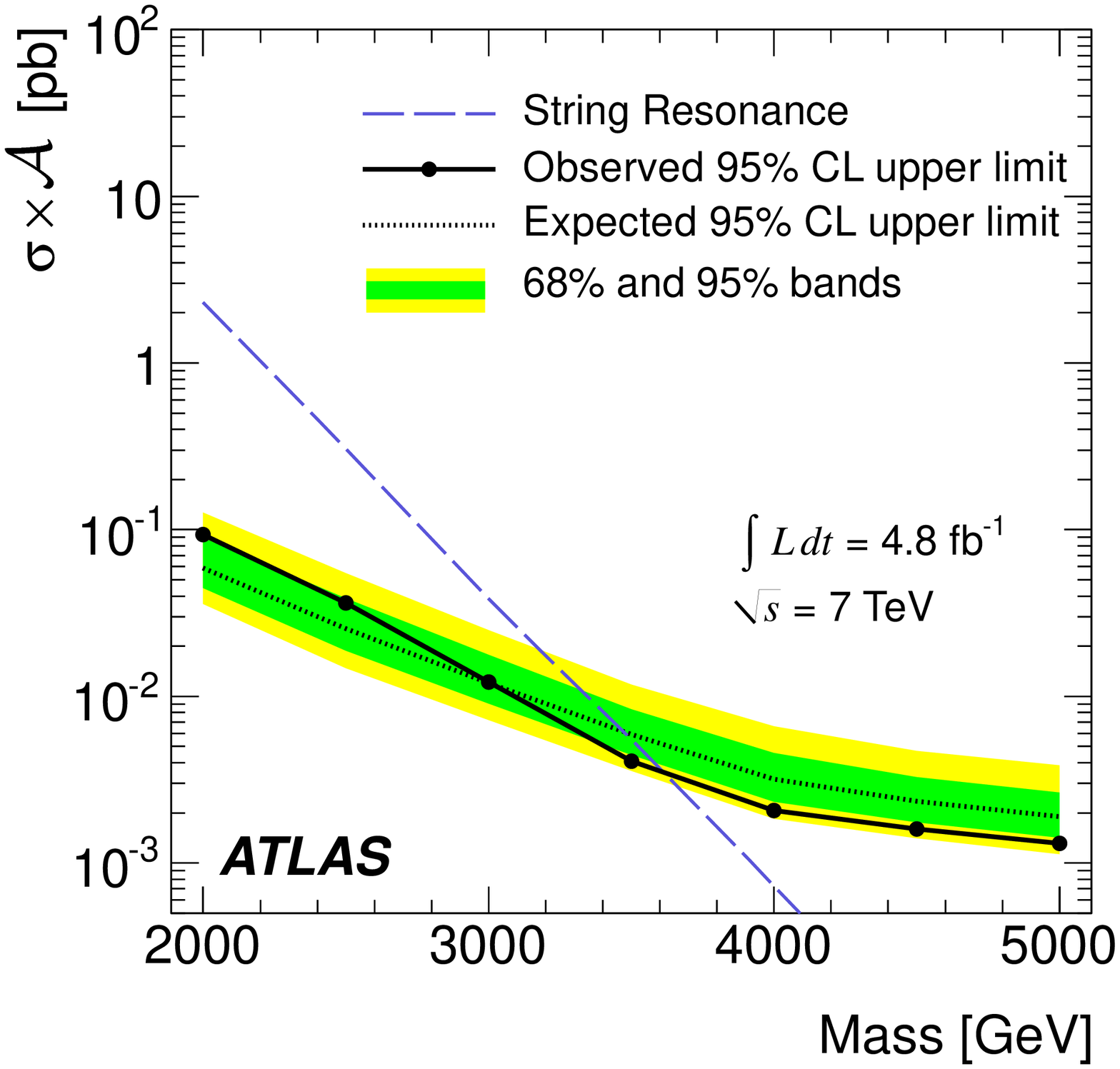}}
    \caption{
      In (a), 95\% CL upper limits on $\sigma\times {\cal A}$
      $\times$ BR as a function of 
      particle mass (black filled circles) from $\mjj$ analysis 
      are shown for heavy gauge bosons, $\Wprime$. The black dotted curve shows 
      the 95\% CL upper limit expected in the absence of any resonance signal, and the 
      green and yellow bands represent the 68\% and 95\% contours
      of the expected limit, respectively. The observed (expected) limit
      occurs at the crossing of the dashed theoretical $\sigma\times {\cal A}$$\times$ 
      BR curve with the observed (expected) 95\% CL upper limit curve.
      In (b), 95\% CL upper limits on $\sigma\times {\cal A}$ are shown
      for string resonances, SR, with the 
      equivalent set of contours for this model, and the same method of
      limit determination.
    }
    \label{fig:combinedfigW}
  \end{figure*}

The limits for heavy charged gauge bosons, $\Wprime$, are shown in figure \ref{fig:limwprimemjj}. 
For this model, only final states with dijets have been simulated. 
The branching ratio, BR, to the studied $q\bar{q}'$ final state varies
little with mass and is 0.75 for $m_{\Wprime}$ values of 1.1~TeV to
3.6~TeV, and the acceptance ranges from 29\% to 36\%. 
The expected mass limit at 95\% CL is \MjjLimitExpectedWprime~TeV,
and the observed limit is \MjjLimitWprime~TeV.
This is the first time that an ATLAS limit on $\Wprime$ production
is set using the dijet mass distribution. Searches for leptonic decays
of the $\Wprime$ are however expected to be more sensitive.

The $\Wprime$ hypothesis used in the current study assumes SM couplings to
quarks.  If a similar model were to predict stronger couplings, for example,
figure \ref{fig:limwprimemjj} could be used to estimate the new
mass limit by shifting the theoretical curve upward by the ratio of
the squared couplings.
Alternately, the current limit on $\Wprime$ decaying to dijets could be
of interest for comparison with leptophobic $\Wprime$ models, where all
final states would be hadronic ~\cite{Georgi1990541,Grojean:2011vu,RHW1,RHW2}.

The limits for string resonances are shown in figure \ref{fig:limsrmjj}. 
The SR acceptance ranges from 45\% to 48\% for masses varying from 2.0~TeV to 5.0~TeV.
The expected mass limit at 95\% CL is
\MjjLimitExpectedStrRes~TeV, and the observed limit is \MjjLimitStrRes~TeV. 

Tables with  acceptance values and limits for all models discussed here
can be found in appendix \ref{sec:tablimits:mjj}.

\section{Model-independent limits on dijet resonance production}
\label{Sec:ModelIndep}

As in previous dijet resonance analyses, limits on dijet resonance production
are determined here using a Gaussian resonance shape hypothesis.
Limits are set for a collection
of hypothetical signals that are assumed to be Gaussian-distributed
in $\mjj$ with means ($m_{\text{G}}$) ranging from 1.0~TeV to 4.0~TeV and with
standard deviations ($\sigma_{\text{G}}$) from 7\% to 15\% of the mean.

Systematic uncertainties are treated using the same methods as applied in the
model-dependent limit setting described above.  
The only difference between the Gaussian analysis and the standard
analysis is that the decay of the dijet final state is not simulated.
In place of this, it is assumed that the dijet signal mass distribution
is Gaussian in shape, and the JES uncertainty is modelled as an uncertainty
of 4\% in the central value of the Gaussian signal.  
This approach has been validated by shifting the energy of all jets in \Pythia~6
signal templates by their JES uncertainty and evaluating the relative shift
of the mass peak.

\begin{figure}[!htb]
  \centering 
  \includegraphics[width=0.70\textwidth]{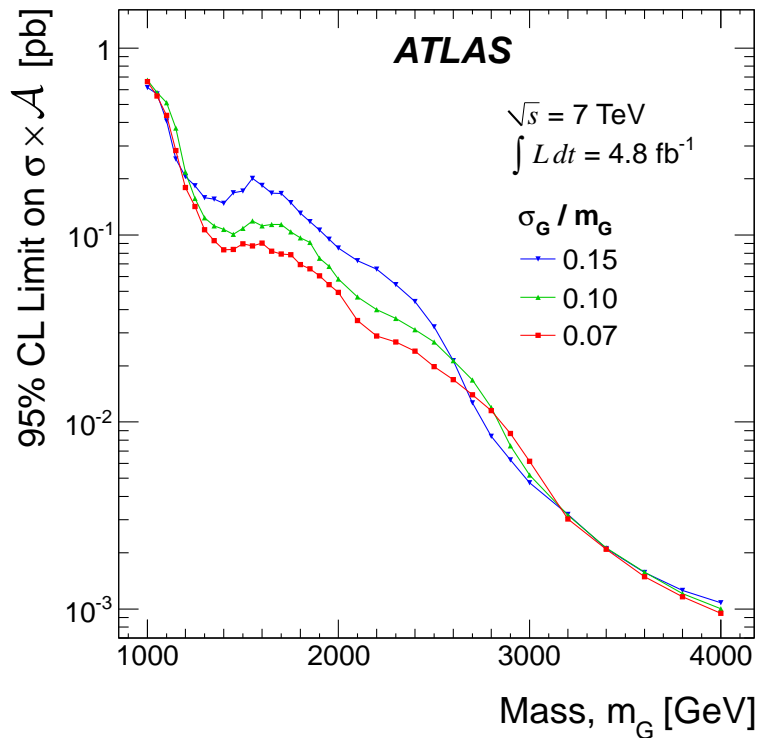}
  \caption{The 95\%\ CL upper limits on $\sigma\times {\cal A}$ for a simple Gaussian
           resonance decaying to dijets as a function of the mean mass, $m_{\text{G}}$, for
           three values of $\sigma_{\text{G}}/m_{\text{G}}$, taking into account
           both statistical and systematic uncertainties.}
  \label{fig:modindeplims}
\end{figure}

The resulting limits on $\sigma\times {\cal A}$ for
the Gaussian template model are shown in figure
\ref{fig:modindeplims} and detailed in table  \ref{qnum.gausslims}.
These results may be utilised to set limits on NP models beyond those
considered in the current studies, under the condition that their
signal shape approaches a Gaussian distribution after applying the
kinematic selection criteria on $y^*$, $\mjj$  and $\eta$ of the
leading jets (section \ref{sec:recocuts}).  The acceptance should
include the branching ratio of the particle decaying into dijets and
the physics selection efficiency. The ATLAS $\mjj$ resolution is about 5\%,
hence NP models with a width smaller than 7\% should
be compared to the 7\% column of table \ref{qnum.gausslims}. 
Models with a greater width should use the column that best matches their width. 
A detailed description of the recommended procedure, including the treatment
of detector resolution effects, is given in ref.~\cite{ATLAS:Res2011}.

\begin{table}[!htb]
\begin{center}
\begin{tabular}{|l|lll|}
\hline
 &         \multicolumn{3}{|c|}{Observed 95\% CL upper limits on $\sigma\times {\cal A}$ [pb]} \\
$m_{\text{G}}$ [GeV]   & $\sigma_{\text{G}}/m_{\text{G}} =$ 7\%   &  $\sigma_{\text{G}}/m_{\text{G}} =$ 10\%  &  $\sigma_{\text{G}}/m_{\text{G}} =$ 15\%  \\
\hline
  1000 &  0.66 &  0.67 &  0.61  \\
  1050 &  0.56 &  0.58 &  0.57  \\
  1100 &  0.44 &  0.51 &  0.41  \\
  1150 &  0.28 &  0.37 &  0.26  \\
  1200 &  0.18 &  0.22 &  0.21  \\
  1250 &  0.14 &  0.16 &  0.18  \\
  1300 &  0.11 &  0.12 &  0.16  \\
  1350 &  0.093 &  0.11 &  0.16  \\
  1400 &  0.083 &  0.11 &  0.15  \\
  1450 &  0.084 &  0.10 &  0.17  \\
  1500 &  0.090 &  0.11 &  0.17  \\
  1550 &  0.087 &  0.12 &  0.20  \\
  1600 &  0.090 &  0.11 &  0.18  \\
  1650 &  0.082 &  0.11 &  0.17  \\
  1700 &  0.079 &  0.11 &  0.17  \\
  1750 &  0.078 &  0.10 &  0.15  \\
  1800 &  0.069 &  0.097 &  0.13  \\
  1850 &  0.066 &  0.091 &  0.12  \\
  1900 &  0.061 &  0.075 &  0.11  \\
  1950 &  0.054 &  0.068 &  0.095  \\
  2000 &  0.049 &  0.058 &  0.085  \\
  2100 &  0.035 &  0.047 &  0.073  \\
  2200 &  0.029 &  0.040 &  0.066  \\
  2300 &  0.027 &  0.036 &  0.054  \\
  2400 &  0.024 &  0.031 &  0.044  \\
  2500 &  0.020 &  0.027 &  0.032  \\
  2600 &  0.017 &  0.021 &  0.021  \\
  2700 &  0.014 &  0.017 &  0.013  \\
  2800 &  0.012 &  0.012 &  0.0084  \\
  2900 &  0.0087 &  0.0075 &  0.0063  \\
  3000 &  0.0062 &  0.0052 &  0.0047  \\
  3200 &  0.0030 &  0.0032 &  0.0032  \\
  3400 &  0.0021 &  0.0021 &  0.0021  \\
  3600 &  0.0015 &  0.0016 &  0.0016  \\
  3800 &  0.0012 &  0.0012 &  0.0013  \\
  4000 &  0.0010 &  0.0010 &  0.0011  \\
\hline
\end{tabular}
\end{center}
\caption{The 95\% CL upper limit on $\sigma\times {\cal A}$ [pb] for the Gaussian model.  
     The symbols $m_{\text{G}}$ and $\sigma_{\text{G}}$ are, respectively,
     the mean mass and standard deviation of the Gaussian.}
\label{qnum.gausslims}
\end{table}

\section{Limits on CI and QBH from the $\chi$ distributions}
\label{section:chiLimit}

The $\chi$ distribution in the highest mass bin of figure \ref{fig:chisvsQCD}
is used to set 95\% CL limits on two NP hypotheses, CI and QBH. 

In the contact interaction analysis, four MC samples of QCD production modified by
a contact interaction are created for values of $\Lambda$\ ranging from 4.0~TeV to 10.0~TeV.  
For the CI distributions, QCD K-factors are applied to the QCD-only component of
the cross section, as follows: before normalising the $\chi$-distributions to unit
area, the LO QCD part of the cross section, determined from a QCD-only simulation
sample, is replaced by the QCD cross section corrected for NLO effects. 

Using the QCD distribution and the finite set of MC CI distributions, each $\chi$-bin is
fit as function of $\Lambda$ against a four-parameter
interpolation function\footnote{The fitting function is
  $f(x) =  p_4  / \,  \left[\mathrm{exp}\left(p_1 \, (p_2 - \mathrm{log} (x) )
 \right)+1\right] +p_3$, $x = 1/\Lambda^2$.},
allowing for a smooth integration of the posterior
probability density functions over $\Lambda$.
From the signal fits, a posterior probability density
is constructed as a function of $\Lambda$.
The systematic uncertainties described in section \ref{sec:compare:chi} are convolved with the posterior distribution through
pseudo-experiments drawn from the NP hypotheses. For the expected limit,
pseudo-experiments are performed on the QCD background and used as pseudo-data.

\begin{figure}[h!]
  \centering 
  \includegraphics[width=0.7\textwidth]{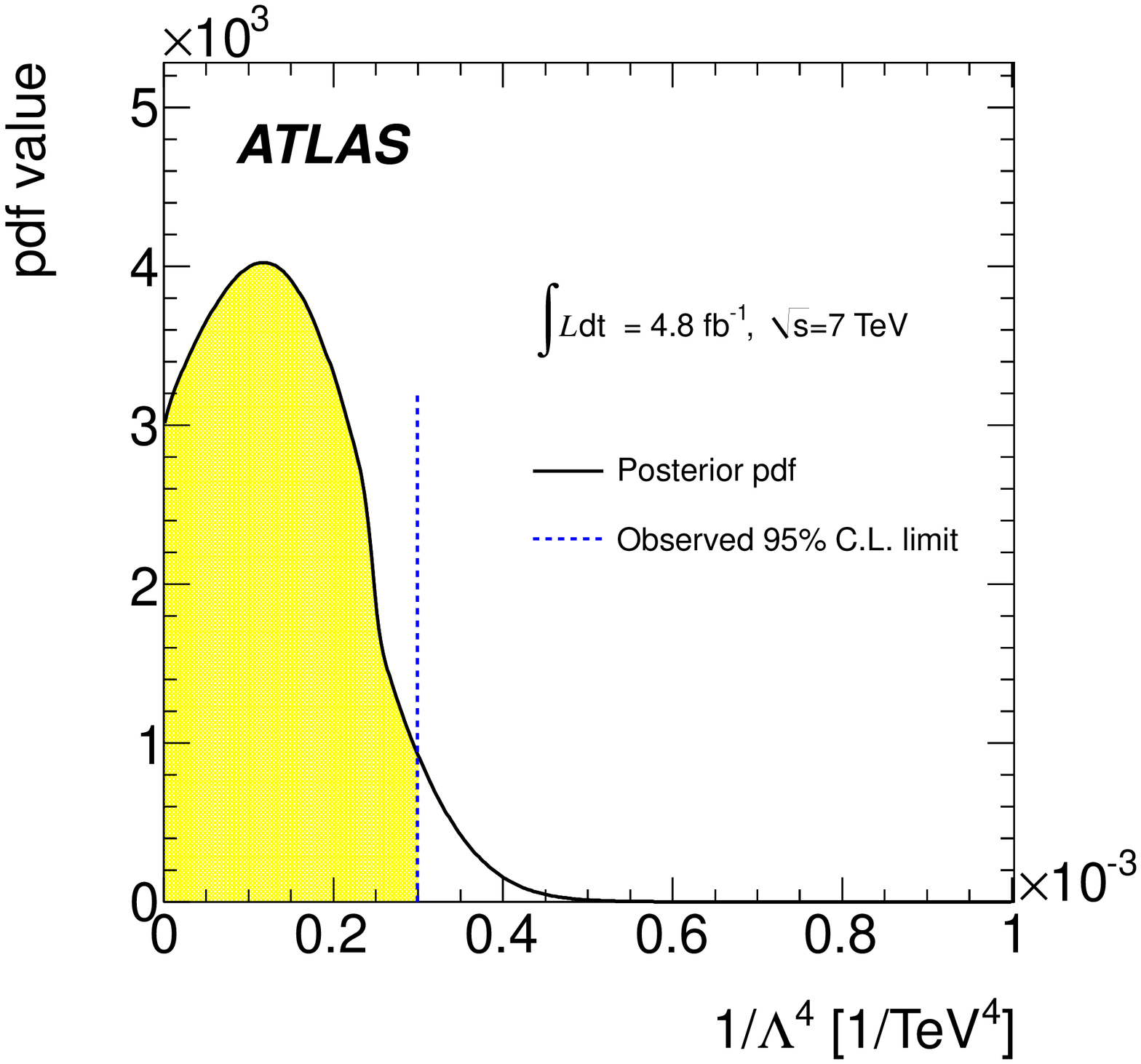}
  \caption{Observed posterior probability density function as
    function of $1/\Lambda^4$ for the CI model. The coloured area
    shows the 95\% area, and the blue dashed line denotes the 95\% CL~limit.} 
  \label{fig:chilimitCI}
\end{figure}

This analysis sets a 95\%\
CL lower limit on $\Lambda $ at \ElevenBinChiLambdaDest~TeV with an expected
limit of \ElevenBinChiExpectedLambdaDest~TeV. The observed posterior
probability density function is shown in figure \ref{fig:chilimitCI}.

To test the sensitivity of the CI limit to the choice of prior, this analysis is repeated for a constant prior in $1/\Lambda^2$, which has
been used in previous publications.  As anticipated, the expected limit is less conservative, increasing by 0.40~TeV.  Since the
constant prior in $1/\Lambda^4$ more accurately follows the cross section predicted for CI, the $1/\Lambda^2$ result is not reported in
the final results of the current studies.

The second model is QBH with $n = 6$ and with a constant prior in $1/M_D^4$,
which is for $n=6$ proportional to the  cross section. 
Similarly to what is done for CI, the QCD sample, together
with a set of eleven QBH samples with $M_D$ ranging from 2.0~TeV
to 6.0~TeV, is fit to the same smooth function in every $\chi$-bin 
to enable integration of the posterior probability density functions
over $M_D$. The expected and observed 95\%\ CL lower limits on $M_D$ are
\ElevenBinChiExpectedQBH~TeV and  \ElevenBinChiQBH~TeV, respectively.

\section{Limits on new resonant phenomena from the \Fchimjj\ distribution}

The Bayesian approach employed to set exclusion limits on new resonant phenomena with
the dijet mass spectrum may be applied to the \Fchimjj\ distribution
(see figure \ref{fig:fchicompQCD}), provided
that the NP models under consideration do not include interference with QCD.
Unlike the $\mjj$ resonance analysis, the background prediction is based on the QCD MC
samples processed through detector simulation and corrected for NLO effects.
The likelihood is constructed from two $\mjj$ distributions and their associated
uncertainties, one distribution being the numerator spectrum of the \Fchimjj\
distribution and the other being the denominator.  Here too, pseudo-experiments
are used to convolve all systematic uncertainties, which in this case include the
JES uncertainties, and the PDF and scale uncertainties associated with the QCD prediction.

\begin{figure}[h!]
  \centering \includegraphics[width=0.7\textwidth]{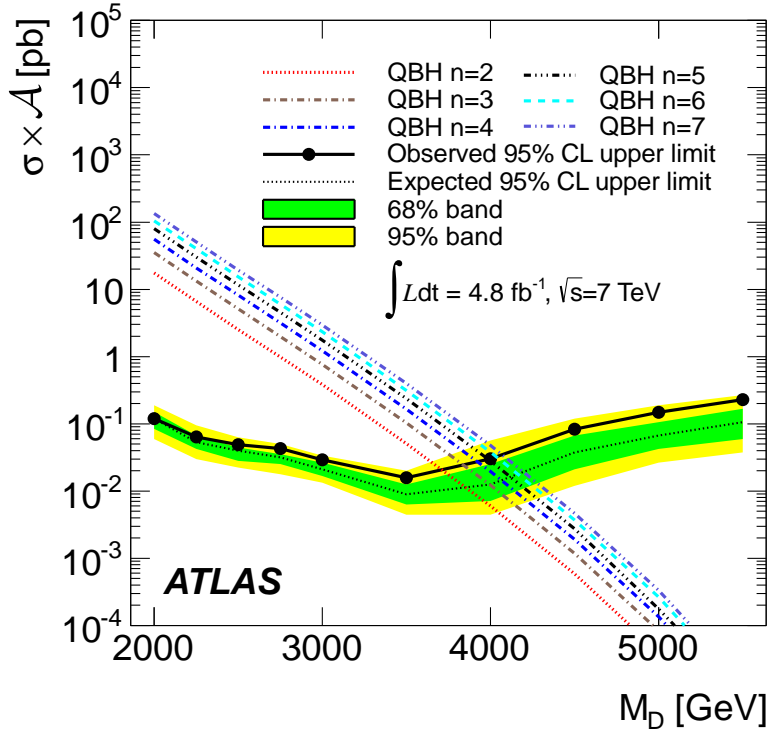}
  \caption{The 95\% CL upper limits on $\sigma\times {\cal A}$ as
    function of the reduced Planck mass $M_D$ of the QBH  model
    using \Fchimjj\
    (black filled circles). The black dotted curve shows the 95\% CL upper limit expected from Monte Carlo, and the 
		green and yellow bands represent the 68\% and 95\% contours
		of the expected limit, respectively. Theoretical
                predictions of $\sigma\times {\cal A}$ are shown for
                various numbers of extra dimensions. } 
  \label{fig:fchimjjbayesqbh}
\end{figure}

Figure \ref{fig:fchimjjbayesqbh} shows the limits expected and observed from data on the production
cross section $\sigma$ times the acceptance $\cal A$, along with theoretical predictions for
the QBH model \cite{RandallMeade,Feng:2004}, for $n$ ranging from two to seven. 
For this model, generator-level studies have shown that the acceptance does not depend on the
number of extra dimensions within this range. Therefore
only the QBH MC sample for $n$ = 6 has been processed through the ATLFAST~2.0 detector simulation, and the acceptance calculated from this sample
is used for all values of $n$. The acceptance is close to 90\% for all $M_D$ values. 
The resulting 95\% CL exclusion limits for the number of extra dimensions $n$
ranging from 2 to 7 are shown in table \ref{tab:qbhlimits}. 

\begin{table}[h!]
\centering {
\begin{tabular}{|l|l|l|}
\hline
$n$ extra    &  Expected   &  Observed    \\ 
dimensions   & limit [TeV] &  limit [TeV] \\ \hline
    2        &     3.85    &     3.71 \\
    3        &     3.99    &     3.84 \\
    4        &     4.07    &     3.92 \\
    5        &     4.12    &     3.99 \\
    6        &    \FchimjjLimitExpectedQBH    &     \FchimjjLimitQBH\ \\
    7        &     4.19    &     4.07 \\
\hline
\end{tabular}
     }
\caption {Lower limits at 95\% CL on $M_D$ of the QBH model with $n=2$
  to 7 extra dimensions.}
\label{tab:qbhlimits}

\end{table}

The same analysis is applied to detect resonances in \Fchimjj\ due to excited quarks.
With an acceptance close to 90\% for all masses this analysis sets a
95\%\ CL lower limit on $m_{q^\ast}$ at \FchimjjLimitQstar~TeV with an expected
limit of \FchimjjLimitExpectedQstar~TeV.

\section{Limits on CI from the \Fchimjj\ distribution}

As was done previously with the ATLAS 2010 data sample~\cite{Aad:2011aj},
the \Fchimjj\ distribution (see figure \ref{fig:fchicompQCD}) is used
in the current study to set limits on quark contact interactions.

The procedure is very similar to  the one used for limits obtained with
$\chi$ discussed in section \ref{section:chiLimit}. 
MC samples of QCD production modified by a contact interaction are
created for values of $\Lambda$\ ranging from 4.0~TeV to 10.0~TeV.  
For the CI distributions, QCD K-factors are applied to the QCD-only
components of the numerator
and denominator of \Fchimjj\ separately. This is done by subtracting the
LO QCD cross section and adding the QCD cross section corrected for NLO effects. 

Simulated \Fchimjj\ distributions are statistically smoothed by a fit in $\mjj$.
For the pure QCD sample (corresponding to $\Lambda = \infty$), a second-order
polynomial is used, while for the MC distributions with finite
$\Lambda$, a Fermi function is added to the polynomial, which gives a
good representation of the onset of contact interactions.

Next, all $\mjj$ bins of the MC \Fchimjj\ distributions are interpolated in $\Lambda$
using the same four-parameter interpolation function  used for the $\chi$
analysis, creating a smooth
predicted
\Fchimjj\ surface as a function of $\mjj$\ and $\Lambda$.  This surface
enables integration in $\mjj$\ vs. $\Lambda$ for continuous values of $\Lambda$.

Pseudo-experiments 
are then employed to construct a posterior
probability, assuming a prior that is flat in $1/\Lambda^4$.
This analysis sets a 95\%\ CL lower limit 
on $\Lambda$ at \FchimjjLimitLambdaDest~TeV with an expected limit
of \FchimjjLimitExpectedLambdaDest~TeV. 


\section{Conclusions}

Dijet mass and angular distributions have been measured by the ATLAS
experiment over a large angular range and spanning dijet masses
up to approximately $\HighestDijetMass$~TeV, using \integLumi\ of $pp$
collision data at $\sqrt{s} = 7$~TeV.
No resonance-like features have been observed in the dijet mass spectrum, 
and all angular distributions are consistent with QCD predictions. 
This analysis places limits on a variety of hypotheses for physics
phenomena beyond the Standard Model, as 
summarised in table \ref{tab:Summary}.

\begin{table}[h!]
\centering {
\begin{tabular}{lcc}
\hline
Model and Analysis Strategy  \quad \quad &  \multicolumn{2}{c}{95\%\ CL Limits [TeV]} \\
            &  Expected  &  Observed\\
\hline \hline \multicolumn{3}{c}{Excited quark, mass of $\qstar$ } \\ \hline
   Resonance in $\mjj$     &   \MjjLimitExpectedqstar           & \MjjLimitqstar  \\    
   Resonance in \Fchimjj\  &   \FchimjjLimitExpectedQstar       & \FchimjjLimitQstar  \\    
\hline \hline \multicolumn{3}{c}{Colour octet scalar, mass of s8 } \\ \hline
   Resonance in $\mjj$     &    \MjjLimitExpecteds8              & \MjjLimits8  \\    
\hline \hline \multicolumn{3}{c}{Heavy $W$ boson, mass of $\Wprime$ } \\ \hline
   Resonance in $\mjj$     &   \MjjLimitExpectedWprime            & \MjjLimitWprime  \\    
\hline \hline \multicolumn{3}{c}{String resonances, scale of SR} \\ \hline
   Resonance in $\mjj$     &   \MjjLimitExpectedStrRes           & \MjjLimitStrRes  \\    
\hline \hline \multicolumn{3}{c}{Quantum Black Hole for $n=6$, $M_D$} \\ \hline
   \Fchimjj\     &   \FchimjjLimitExpectedQBH         &  \FchimjjLimitQBH  \\    
   $\chi$\ , $\mjj > 2.6$~TeV &
                               \ElevenBinChiExpectedQBH         &  \ElevenBinChiQBH \\
\hline \hline \multicolumn{3}{c}{ Contact interaction, $\Lambda$, destructive interference} \\ \hline
   \Fchimjj       &    \FchimjjLimitExpectedLambdaDest     &  \FchimjjLimitLambdaDest   \\
   $\chi$\ , $\mjj > 2.6$~TeV 
                           &    \ElevenBinChiExpectedLambdaDest     &  \ElevenBinChiLambdaDest \\
\hline
\end{tabular}
}     
\caption {The 95\% CL lower limits on the masses and energy scales of the models 
examined in this study.  
All limit analyses are Bayesian, with statistical and systematic uncertainties 
included. For each NP hypothesis, the result corresponding to the highest
expected limit is the result quoted in the abstract.
}
\label{tab:Summary}
\end{table}

\begin{table}[!htb]
\begin{center}
\begin{tabular}{l|lll}
\hline 
New Phenomenon &  36 pb$^{-1}$~\cite{Aad:2011aj}  &   1.0 fb$^{-1}$~\cite{ATLAS:Res2011} 
 & \integLumi current\\
\hline 
 \multicolumn{4}{c}{Resonance in $\mjj$} \\
\hline
  Excited quark, mass of $\qstar$     &  2.07  & 2.81   & \MjjLimitExpectedqstar  \\ 
 Colour octet scalar, mass of s8          &  -    & 1.77  & \MjjLimitExpecteds8\\ 
\hline \multicolumn{4}{c}{Angular distribution in $\chi$} \\
\hline
 Contact interaction, $\Lambda$ &   5.4   &   -   &   \ElevenBinChiExpectedLambdaDest \\ 
\hline
\end{tabular}
\end{center}
\caption{ATLAS previous and current expected 95\% CL upper limits
  [TeV] on new phenomena. The current expected limit for $\qstar$ cannot be compared directly
to the two previous limits since they employed \Pythia~6 samples with
an error in the simulation of final state radiation.  Had such samples
been used in the current analysis, the expected $\qstar$ limit would be
0.10~TeV higher.}
\label{tab:complims}
\end{table}

For $\sqrt{s} = 7$~TeV $pp$ collisions at the LHC, the integrated luminosity
used in the current studies represents a substantial increase over
that available in previously published ATLAS dijet searches. 
Table~\ref{tab:complims} lists the previous and current expected limits from
ATLAS studies using dijet analyses for three benchmark models:
excited quarks, colour octet scalars, and contact interactions with
destructive interference.
The increase in the excited quark mass limit would have been greater by 0.10~TeV
had there not been the long-standing problem with the default \Pythia~6
$\qstar$ model, discussed in earlier sections.

For 2012 running, the collision energy of the LHC has been raised from 7~TeV
to 8~TeV.  The higher energy, and the associated rise in parton 
luminosity, will increase search sensitivities and the
possibility of discoveries.
The current 2011 analysis provides a reference for the study of
energy-dependent effects once the 2012 data set has been analysed.



\clearpage

\acknowledgments
We thank Noriaki Kitazawa for the string resonance amplitude calculations and event samples. 

We thank CERN for the very successful operation of the LHC, as well as the
support staff from our institutions without whom ATLAS could not be
operated efficiently.

We acknowledge the support of ANPCyT, Argentina; YerPhI, Armenia; ARC,
Australia; BMWF and FWF, Austria; ANAS, Azerbaijan; SSTC, Belarus; CNPq and FAPESP,
Brazil; NSERC, NRC and CFI, Canada; CERN; CONICYT, Chile; CAS, MOST and NSFC,
China; COLCIENCIAS, Colombia; MSMT CR, MPO CR and VSC CR, Czech Republic;
DNRF, DNSRC and Lundbeck Foundation, Denmark; EPLANET, ERC and NSRF, European Union;
IN2P3-CNRS, CEA-DSM/IRFU, France; GNSF, Georgia; BMBF, DFG, HGF, MPG and AvH
Foundation, Germany; GSRT and NSRF, Greece; ISF, MINERVA, GIF, DIP and Benoziyo Center,
Israel; INFN, Italy; MEXT and JSPS, Japan; CNRST, Morocco; FOM and NWO,
Netherlands; BRF and RCN, Norway; MNiSW, Poland; GRICES and FCT, Portugal; MERYS
(MECTS), Romania; MES of Russia and ROSATOM, Russian Federation; JINR; MSTD,
Serbia; MSSR, Slovakia; ARRS and MVZT, Slovenia; DST/NRF, South Africa;
MICINN, Spain; SRC and Wallenberg Foundation, Sweden; SER, SNSF and Cantons of
Bern and Geneva, Switzerland; NSC, Taiwan; TAEK, Turkey; STFC, the Royal
Society and Leverhulme Trust, United Kingdom; DOE and NSF, United States of
America.

The crucial computing support from all WLCG partners is acknowledged
gratefully, in particular from CERN and the ATLAS Tier-1 facilities at
TRIUMF (Canada), NDGF (Denmark, Norway, Sweden), CC-IN2P3 (France),
KIT/GridKA (Germany), INFN-CNAF (Italy), NL-T1 (Netherlands), PIC (Spain),
ASGC (Taiwan), RAL (UK) and BNL (USA) and in the Tier-2 facilities
worldwide.


\clearpage
\bibliographystyle{JHEP}
\bibliography{ExoticDijetsJHEP2012}

\appendix
\newpage
\section{ Limits on new resonant phenomena from the $\mjj$ distribution}
\label{sec:tablimits:mjj}
\subsection{Excited quarks}


\begin{table}[!htb]
\begin{center}
\begin{tabular}{|l|l|l|l|l|l|}
\hline
 $m_{\qstar}$ $[\mathrm{GeV}]$ & Observed         &  Expected & Expected $\pm 1
 \sigma $ &   Expected $\pm 2  \sigma $  & $\cal A$  \\
\hline
1000 & 1.43 & 0.55 & 0.36/1.064 & 0.31/1.58  &0.299\\
1200 & 0.30 & 0.36 & 0.27/0.66 & 0.23/0.99 &0.403\\
1400 & 0.16 & 0.22 & 0.17/0.35 & 0.14/0.52  &0.459 \\
1600 & 0.16 & 0.15 & 0.12/0.25 & 0.098/0.37 &0.481\\
1800 & 0.16 & 0.10 & 0.079/0.16 & 0.065/0.24 &0.497\\
2000 & 0.12 & 0.071 & 0.054/0.11 & 0.043/0.16 &0.501 \\
2250 & 0.064 & 0.045 & 0.034/0.070 & 0.027/0.10 &0.505\\
2500 & 0.050 & 0.032 & 0.023/0.050 & 0.018/0.071 &0.511 \\
2750 & 0.032 & 0.023 & 0.016/0.036 & 0.013/0.051 &0.499 \\
3000 & 0.017 & 0.016 & 0.012/0.024 & 0.0094/0.034 &0.500 \\
3250 & 0.0081 & 0.011 & 0.0086/0.017 & 0.0069/0.024 &0.505\\
3500 & 0.0056 & 0.0081 & 0.0062/0.012 & 0.0049/0.016  &0.499\\
3750 & 0.0041 & 0.0063 & 0.0047/0.0090 & 0.0037/0.013 &0.493 \\
4000 & 0.0034 & 0.0049 & 0.0036/0.0070 & 0.0028/0.010 &0.484\\
\hline
\end{tabular}
\end{center} 
\caption{The 95\% CL upper limit on $\sigma\times {\cal A}$ [pb] for excited quarks, $\qstar$.}
\end{table}

\subsection{Colour octet scalars}

\begin{table}[!htb]
\begin{center}
\begin{tabular}{|l|l|l|l|l|l|}
\hline
 $m_{\textrm{s8}}$ $[\mathrm{GeV}]$ & Observed         &  Expected & Expected $\pm 1 \sigma $ &   Expected $\pm 2  \sigma $   & $\cal A$ \\
\hline

1300 & 0.40 & 0.68 & 0.38/1.45 & 0.31/2.20 & 0.339\\
1500 & 0.27 & 0.38 & 0.27/0.75 & 0.23/1.18 & 0.405\\
1700 & 0.24 & 0.27 & 0.20/0.52 & 0.17/0.79 &  0.443\\
2000 & 0.33 & 0.16 & 0.12/0.29 & 0.099/0.43 & 0.467 \\
2500 & 0.17 & 0.084 & 0.059/0.14 & 0.049/0.21 &  0.484\\
3000 & 0.097 & 0.062 & 0.042/0.11 & 0.034/0.17  & 0.441\\
3500 & 0.034 & 0.049 & 0.036/0.079 & 0.030/0.12 & 0.390\\
4000 & 0.035 & 0.048 & 0.038/0.073 & 0.032/0.11 & 0.357 \\


\hline
\end{tabular}
\end{center} 
\caption{The 95\% CL upper limit on $\sigma\times {\cal A}$ [pb] for
  colour octets scalars, s8.}
\end{table}

\clearpage
\subsection{Heavy $W$ boson}

\begin{table}[!htb]
\begin{center}
\begin{tabular}{|l|l|l|l|l|l|}
\hline
 $m_{\Wprime}$ $[\mathrm{GeV}]$ & Observed         &  Expected & Expected $\pm 1 \sigma $ &   Expected $\pm 2  \sigma $   & $\cal A$ \\
\hline
1100 & 0.65 & 0.46 & 0.32/0.88 & 0.27/1.30 & 0.286 \\
1200 & 0.29 & 0.35 & 0.26/0.62 & 0.22/0.90 & 0.314  \\
1400 & 0.15 & 0.21 & 0.16/0.33 & 0.13/0.48 & 0.345\\
1600 & 0.15 & 0.14 & 0.11/0.23 & 0.094/0.33 & 0.358\\
1800 & 0.13 & 0.099 & 0.077/0.16 & 0.063/0.23 & 0.353 \\
2000 & 0.12 & 0.072 & 0.055/0.11 & 0.045/0.16 & 0.341 \\
2400 & 0.065 & 0.042 & 0.031/0.064 & 0.025/0.090 & 0.293\\

\hline
\end{tabular}
\end{center} 
\caption{The 95\% CL upper limit on $\sigma\times {\cal A}$ $\times$ BR [pb] for
 Heavy $W$ bosons, $\Wprime$.}
\end{table}

\subsection{String resonances}

\begin{table}[!htb]
\begin{center}
\begin{tabular}{|l|l|l|l|l|l|}
\hline
 $m_{\textrm{SR}}$ $[\mathrm{GeV}]$ & Observed         &  Expected & Expected $\pm 1 \sigma $ &   Expected $\pm 2  \sigma $  & $\cal A$  \\
\hline
2000 & 0.094 & 0.059 & 0.041/0.080 & 0.032/0.12  & 0.449  \\
2500 & 0.036 & 0.026 & 0.017/0.034 & 0.013/0.048 & 0.447 \\
3000 & 0.012 & 0.012 & 0.0077/0.016 & 0.0061/0.022 & 0.452 \\
3500 & 0.0041 & 0.0059 & 0.0036/0.0069 & 0.0028/0.010 & 0.464\\
4000 & 0.0021 & 0.0032 & 0.0020/0.0038 & 0.0016/0.0058 & 0.458\\
4500 & 0.0016 & 0.0023 & 0.0016/0.0029 & 0.0013/0.0040 & 0.478\\
5000 & 0.0013 & 0.0019 & 0.0012/0.0024 & 0.0010/0.0034 & 0.482 \\
\hline
\end{tabular}
\end{center} 
\caption{The 95\% CL upper limit on $\sigma\times {\cal A}$ [pb] for
 string resonances, SR.}
\end{table}

\clearpage
\begin{flushleft}
{\Large The ATLAS Collaboration}

\bigskip

G.~Aad$^{\rm 48}$,
T.~Abajyan$^{\rm 21}$,
B.~Abbott$^{\rm 111}$,
J.~Abdallah$^{\rm 12}$,
S.~Abdel~Khalek$^{\rm 115}$,
A.A.~Abdelalim$^{\rm 49}$,
O.~Abdinov$^{\rm 11}$,
R.~Aben$^{\rm 105}$,
B.~Abi$^{\rm 112}$,
M.~Abolins$^{\rm 88}$
% ...
\bigskip
\\
% ...
$^{11}$ Institute of Physics, Azerbaijan Academy of Sciences, Baku, Azerbaijan\\
$^{12}$ Institut de F{\'\i}sica d'Altes Energies and Departament de F{\'\i}sica de la Universitat Aut{\`o}noma de Barcelona and ICREA, Barcelona, Spain\\
% ...
$^{21}$ Physikalisches Institut, University of Bonn, Bonn, Germany\\
% ...
$^{48}$ Fakult{\"a}t f{\"u}r Mathematik und Physik, Albert-Ludwigs-Universit{\"a}t, Freiburg, Germany\\
$^{49}$ Section de Physique, Universit{\'e} de Gen{\`e}ve, Geneva, Switzerland\\
% ...
$^{88}$ Department of Physics and Astronomy, Michigan State University, East Lansing MI, United States of America\\
% ...
$^{105}$ Nikhef National Institute for Subatomic Physics and University of Amsterdam, Amsterdam, Netherlands\\
% ...
$^{111}$ Homer L. Dodge Department of Physics and Astronomy, University of Oklahoma, Norman OK, United States of America\\
$^{112}$ Department of Physics, Oklahoma State University, Stillwater OK, United States of America\\
% ...
$^{115}$ LAL, Universit{\'e} Paris-Sud and CNRS/IN2P3, Orsay, France\\
% ...
\end{flushleft}


\end{document}